\documentclass[sigconf]{acmart}

\copyrightyear{2026}
\acmYear{2026}
\setcopyright{cc}
\setcctype{by}
\acmConference[CHI '26]{Proceedings of the 2026 CHI Conference on Human Factors in Computing Systems}{April 13--17, 2026}{Barcelona, Spain}
\acmBooktitle{Proceedings of the 2026 CHI Conference on Human Factors in Computing Systems (CHI '26), April 13--17, 2026, Barcelona, Spain}
\acmPrice{}
\acmDOI{10.1145/3772318.3791569}
\acmISBN{979-8-4007-2278-3/2026/04}

\usepackage{array}
\usepackage{graphicx}
\usepackage{hyperref}

\begin{document}

\title[Feed Elicitation]{Social Media Feed Elicitation}

\author{Lindsay Popowski}
\email{popowski@stanford.edu}
\orcid{1234-5678-9012}
\affiliation{%
  \institution{Stanford University}
  \city{Palo Alto}
  \state{California}
  \country{USA}
}

\author{Xiyuan Wu}
\email{xiyuan26@stanford.edu}
\orcid{1234-5678-9012}
\affiliation{%
  \institution{Stanford University}
  \city{Palo Alto}
  \state{California}
  \country{USA}
}

\author{Charlotte Zhu}
\email{xuyang1@stanford.edu}
\orcid{1234-5678-9012}
\affiliation{%
  \institution{Stanford University}
  \city{Palo Alto}
  \state{California}
  \country{USA}
}

\author{Tiziano Piccardi}
\email{piccardi@stanford.edu}
\orcid{1234-5678-9012}
\affiliation{%
  \institution{Stanford University}
  \city{Palo Alto}
  \state{California}
  \country{USA}
}

\author{Michael S. Bernstein}
\email{mbernst@stanford.edu}
\orcid{1234-5678-9012}
\affiliation{%
  \institution{Stanford University}
  \city{Palo Alto}
  \state{California}
  \country{USA}
}

\renewcommand{\shortauthors}{Popowski et al.}

\begin{abstract}
  Social media users have repeatedly advocated for control over the currently opaque operations of feed algorithms. Large language models (LLMs) now offer the promise of custom-defined feeds---but users often fail to foresee the gaps and edge cases in how they define their custom feed.
We introduce \textit{feed elicitation interviews}, an interactive method that guides users through identifying these gaps and articulating their preferences to better author custom social media feeds. We deploy this approach in an online study to create custom BlueSky feeds and find that participants significantly prefer the feeds produced from their elicited preferences to those produced by users manually describing their feeds. Through feed elicitation interviews, we advance users' ability to control their social media experience, empowering them to describe and implement their desired feeds.
\end{abstract}


\begin{teaserfigure}
  \centering
    \includegraphics[width=\textwidth]{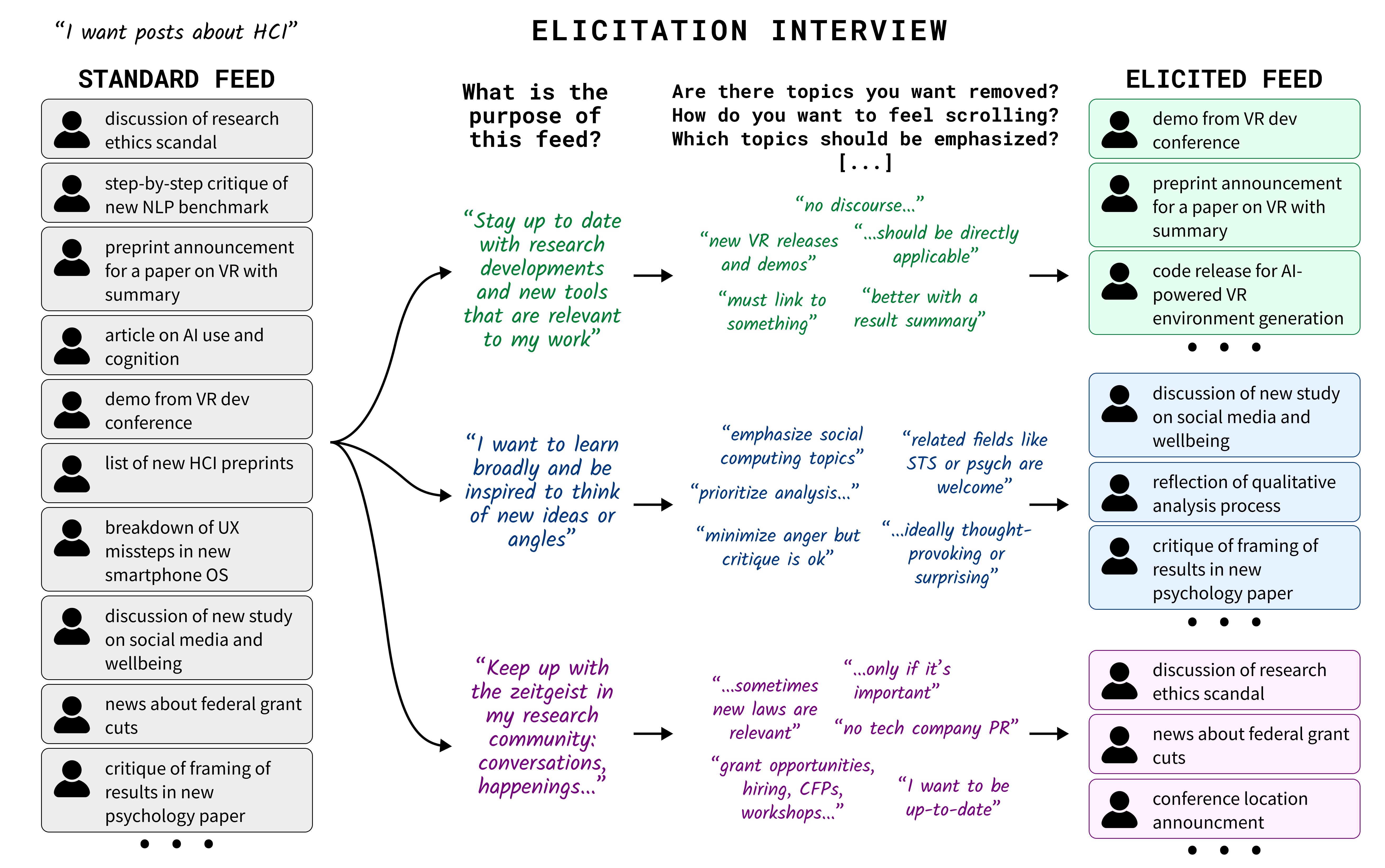}
  \caption{Our feed elicitation interview process guides users through specifying a feed tailored to their ideal use-case and desires.}
  \label{fig: splash}
\end{teaserfigure}
\maketitle

\section{Introduction}

For years, users have bemoaned the tyranny of platform-enforced, engagement-optimizing AI feed algorithms that control the content they see on social media~\cite{bakshy2015exposure,campana2017recommender,cinelli2021echo}. It has been widely speculated that these designs are responsible for many of the ills blamed on social media more broadly: political polarization~\cite{rathje2021out,kubin2021role}; anxiety, depression, and attention issues~\cite{seabrook2016social,allcott2020welfare,shakya2017association,ivie2020meta,shannon2022problematic}; emotional dysregulation~\cite{hormes2014craving}; misinformation~\cite{lazer2018science,valenzuela2019paradox,allcott2017social}; and social disconnection~\cite{bonsaksen2023associations,primack2017social,hunt2018no}. New platforms such as BlueSky are opening the door to a marketplace of different feed algorithms. The questions have now become: what might we want these feeds to accomplish? And, how should the tools to build them work?

In response to this consensus towards promoting user-agency, there have been extensive tool-building efforts to support feed customization and other end-user content-control mechanisms.
Platform-supported features and many popular tools approach this from the moderation space: allowing keyword and user-based blocking, filtering out hateful speech and harassment, crowdsourcing content moderation for harassment~\cite{mahar2018squadbox}, and creating a dashboard to sort content based on career utility~\cite{han2024pressprotect}. 
Researchers have worked on exploring new ranking objectives: in hope of reducing polarization, feed interventions have been run by reverting back to a chronological feed~\cite{guess2023facebookstudy} as well as by downranking partisan animosity~\cite{jialam2024values, piccardi2024algorithms}. Human values have been proposed as a new plausible prescriptive vision for objectives, where feeds are ranked according to how they exhibit values like authenticity, wisdom, or compassion~\cite{kolluri2025alexandria}. 
Finally, a more recent set of tools are allowing users to project their prescriptive visions for feed content. In response to new open API access for feed creation on BlueSky, myriad feed construction tools have emerged with block-based feed logic~\cite{skyfeed2025} and even supporting LLM classification for some attributes and topics~\cite{graze2025}. 

Given the broad technical feasibility today of creating custom feed algorithms, why are they not more widespread? In this paper, we demonstrate an \textit{articulation gap} that is a major bottleneck: even though people strongly desire custom feeds, when they try to construct those feeds, they struggle to foresee and manage all the key choices that are relevant to their desired feed, like whether to promote objective framings on a feed about the news, what level of sarcasm is allowed in a feed for lighthearted humor, or what level of interdisciplinary cross-pollination is desired in a feed to keep up to date on recent research. Without the ability to foresee and handle these decisions, users may be finding the resulting feeds too noisy for practical use. This situation would be predicted by classic psychological research in planning: (1)~the illusion of explanatory depth, where we believe we understand something well only until we must explain it step by step~\cite{rozenblit2002illusion}, and (2)~Kahneman and Tversky's planning fallacy~\cite{kahneman1977intuitive}, where we often bias our planning toward ideal scenarios and fail to foresee edge cases and failure modes. These difficulties are not limited to feed construction: users struggle to fully communicate their wants and needs across a variety of contexts: from information-seeking~\cite{katz1969introduction, brown2008reference} to design~\cite{mohedas2022designinterview} to decision making~\cite{pommeranz2010user}. In the social media context, this means that we produce many half-baked feeds that do not sufficiently satisfy users' desires so as to be useful. 

To close this gap, we propose \textit{feed elicitation interviews}, a method to scaffold users through articulating their preferences and unrealized decisions in authoring a social media feed. This elicitation method uses an interview to force users to actively consider the design space---identifying key choices and unforeseen edge cases of their desired feed---by leveraging a large language model (LLM)'s expansive knowledge of prototypical social media posts that might need to be handled by the algorithm. 
Our LLM-backed interviewing agent questions the user in three main stages to (1) scope the use case of their feed, (2) identify relevant topics, and (3) uncover what post attributes are useful versus detrimental for the feed. In doing so, the interviewer scaffolds articulation of specific preference information that would otherwise be frustrating or demanding for a user to fully dictate (e.g., by suggesting different directions and offering examples), and directs the user's attention towards considerations that they might otherwise overlook (e.g., by presenting unaddressed desiderata).
We demonstrate this approach by embedding it into a feed algorithm construction system that allows users to create custom feeds from BlueSky posts.

We evaluate this feed elicitation approach with a between-subjects study ($N=400$). Participants first complete a manual feed description process to describe their desired feed algorithm (e.g., a feed that identifies discussion of human-computer interaction research). They are then randomized into either a feed elicitation interview, or an interview-less ablation where they receive similar instructions to the interviewer LLM agent but must complete the feed description manually. The result of this process is a textual specification of the feed that our system uses for ranking and relevance judgments over a stream of real BlueSky posts, producing a feed filtered and ranked according to their preferences.

We find that users tended to prefer the feeds generated through this interview process, preferring the interview feeds far more than holding neutral views or preferring the reverse (48\% preferred, 23\% neutral, 29\% disfavorable). In contrast, preferences were split almost evenly between the baseline and ablation conditions. Users also prefer the content in feeds specified via the feed elicitation interview, rating both more posts as desired and fewer as undesired in comparison to the baseline. Again, the ablation condition saw no significant improvement.
Participants had the strongest preference for our elicited feeds when they initially underspecified their specific or unique preferences, and noted enjoying how personalized the feed posts were to their preferences. We do not find any significant cognitive burden or frustration experienced by users through this process.

We contribute identification of the articulation gap as a main barrier of end-user feed algorithms today, a novel LLM-powered elicitation approach that proactively prompts users with refinements and edge cases, and empirical evidence that such elicitation improves the overall preference level as well as the quality of content for the resulting feeds. Ultimately, we argue that when people are scaffolded through the unforeseen decisions at play in feed algorithm design, rather than unknowingly leaving them unconsidered, they are able to produce artifacts that better align with their actual desires.

\section{Related Work}

\subsection{Discontent and Harms Related to Social Media Feeds}

Critiques of social media feeds, especially those driven by modern-day engagement-optimizing algorithms, are widespread and multiplicitous. Issues ranging from political polarization~\cite{rathje2021out,kubin2021role} to body image struggles in young girls~\cite{fardouly2015social,holland2016systematic,saiphoo2019meta} to harassment of marginalized users~\cite{are2020instagram, munn2020angry} to emotional distress in children and adolescents~\cite{wartberg2021relevance} are traced back to the decision of what content people are shown on social media platforms and in what order. 

These critiques are not to be dismissed as spurious, either. With respect to political polarization, a whole range of studies have explored the connection between social media use and partisan animosity~\cite{piccardi2024algorithms, barnidge2017exposure, levy2021social, bail2018exposure}. Further work has attempted to uncover the exact mechanism, and isolate or even reverse the believed polarizing effect of algorithms~\cite{guess2023facebookstudy}. While the exact role of algorithms is not yet uncovered, there are plenty of plausible mechanisms for their impact. For one, because most of these algorithms optimize for engagement or time on platform, they often encourage negative engagement and emotions like anger or upset~\cite{munn2020angry}. When the algorithm amplifies hostile content from the out-group, this increases negative interpartisan sentiment~\cite{milli2025engagement}.

Even when these algorithms work ``well'' (by which we mean, in basic accordance with the priority of upranking content that people like), the downstream effects of these decisions can be detrimental. 
The high-stimulation content often driven towards children on these platforms has been speculated to produce emotional dysregulation~\cite{coyne2021tantrums}. Similarly, the mix of content on broad platforms like TikTok or Twitter or Facebook, where desired content is intermittent, facilitates reward patterns of dopamine which mirror those in gambling to produce addiction~\cite{voinea2024digital}. Even the confusion around algorithmic visibility alone causes stress and angst for social media users~\cite{williams2025why, ma2022difference}.

\subsection{User Feed Controls}

Beyond all of these possible negative side-effects of feeds driven by opaque machine learning algorithms, users are also displeased with them on principle, resenting the opacity and lack of control~\cite{devito2017rip}. Users overwhelmingly desire greater agency in interacting with content on social media, and search for ways to exert agency even through the nondeterministic and noisy channels of content interactions~\cite{eslami2016like, milton2023seeme, hsu2020awareness, stray2024building}. 

Today, most of the tools that allow people to control their social media feed experiences are on the moderation side: removing harmful content. Platforms have their own internal processes to identify and quickly remove content that violates their Terms of Services, as well as extensive reporting flows so users may help identify more. In online communities like those on Reddit or Discord, community moderators also help filter content that may violate more specific rules, tailored to the particular community context and needs, like preventing self-promotion, requiring a certain post format, or banning toxic attitudes~\cite{reddy2023rules, leibmann2025redditrules, seering2024chillbot}. Some infrastructure supports these moderators in creating and enforcing custom rules, on-demand~\cite{chandrasekharan2019crossmod}. Beyond these platform- or community-side endeavors, there are also more approaches that handle moderation on the user side, to assess the validity of potential misinformation~\cite{jahanbakhsh2022peer, jahanbakhsh2024browser}, downrank posts demonstrating partisan animosity to lower affective polarization~\cite{jialam2024values, piccardi2024algorithms}, enlist friends to filter harassment~\cite{mahar2018squadbox}, help journalists sort out harassment and spam~\cite{han2024pressprotect}, and build personalized moderation classifiers~\cite{wang2025authoring}.

The bulk of tools to control social media content have sought to improve user experience by restricting and moderating content on top of the content the platform normally serves to the user. However, in recent years, there has been a surge of feed-control systems to enable new types of feed ranking algorithms. One direction of interest has been towards incorporating human values into social media algorithms~\cite{stray2024building, bernstein2023embedding}, for example by building a set of value primitives that users can combine to create custom feed ranking algorithms~\cite{kolluri2025alexandria}. Another direction involves weaving together multiple sources of content according to user interests, both within a platform~\cite{liu2025decentralized} and across different ones~\cite{bhargava2019gobo}. On BlueSky in particular (which offers the necessary API access to enable end-user feed creation and sharing), tools like Skyfeed~\cite{skyfeed2025} and Graze~\cite{graze2025} have surfaced to allow custom feed algorithm construction, with primitives like topic, keywords, emotional signals, and authorship. Through these tools, tens of thousands of custom feeds are now available for public adoption, some with tens of thousands of users~\cite{blueskyfeeds2025}. Emerging directions allow more comprehensive controls with users dictating the types and sources of posts they want to see at a high level of specificity~\cite{malki2025bonsai}.

\subsection{Design Fixation and the Gulf of Envisioning}

While tooling for social media feeds is starting to facilitate the sorts of control we (and users) have hoped for, gaps remain in the process of feed design that precedes creation. Before a feed is created, users must imagine and articulate what they want, itself a form of design. Whereas toolkits bridge the gulf of execution~\cite{norman1986cognitive} for social media feed creation, there is another gulf in deciding what that feed should be. Prior work has termed this the ``gulf of envisioning,'' which describes how people translate their intentions or desires into the execution or input for execution; for example, translating a desire for output into an LLM prompt~\cite{subramonyam2024gulf}.

One major barrier to envisioning is design fixation, where people narrowly focus in on a set of designs to the extent that they are unable to imagine plausible and useful alternatives~\cite{jansson1991fixation}. Most of users' folk theories for existing feed algorithms coalesce to a handful of main features: engagement features, topical interests, and keywords or toxic content that gets moderated by the platform~\cite{eslami2015assumed, eslami2016like, devito2017rip}.
It would be reasonable to guess that these themes would  color users' expectations and imaginations of feed algorithms.

There are myriad approaches for overcoming design fixation, many of which attempt to help the user reframe their problem space~\cite{moreno2016overcoming}, most commonly via some sort of analogical process involving a parallel problem domain~\cite{moreno2015step}. Our interview process starts by focusing on what purpose the feed is supposed to serve for the user, and lets that drive questions around feed preference. The user is asked to reframe their initial, content- and topic-driven, perspective in order to focus on what the feed should help them accomplish. This reframe helps hone in on potential attributes that are important for a users' given context but may not be generally relevant. However, fixation is not the only reason that users would fail to completely describe their ideal feeds: design does not just require creativity and thinking outside the box, but effective anticipation of and reaction to potential problems.

\subsection{Underspecification and Cognitive Gaps in Planning}

Describing an ideal social media feed is not a simple process of communicating what one has already decided, but rather a complex reasoning process that requires grounding, prioritization, and reflection. Users' initial thoughts are likely to be vague and unstable: fragmented preferences rather than a synthesized set of instructions for the idealized functioning of a feed. In order to produce one’s desired feed, a specification must not only capture abstract intentions but also anticipate the kinds of content it will be applied to.
Designing a social media feed is therefore just one version of a ``wicked problem'' that resists one-shot specification; it requires thoughtful iteration to fully specify.

Design theory makes clear why specifying a feed cannot be treated as a one-time act of planning. Design operates on a reflection-in-action~\cite{schon1986reflective} basis; a designer makes adjustments to their existing perspective on the fly, in reaction to unanticipated stimuli~\cite{schon1986reflective}. Suchman argues that people rarely plan out detailed scripts for future action but instead act provisionally, revising and abandoning initial plans to adapt to the current context~\cite{suchman2007plans}. Similarly, users designing (or planning how to realize) their ideal feeds would need to adapt their feed description when confronted with unexpected posts or new ideas of what content exists, perhaps choosing to remove certain undesirable content or emphasizing characteristics they realize to be essential. In practice, preferences are often only illuminated when a user encounters a stimulus, like a post they feel strong about, that forces them to refine or revise their description. An interactive and reflective process of making and revising would therefore yield a feed very closely tailored to a users' preferences if allowed to continue indefinitely and consume arbitrary time and resources. Our aim is to preempt such a costly direct iteration process by creating an elicitation interview process which serves as an abstracted form of prototyping, surfacing and refining preferences prior to actual execution of the design. 

Psychological research further underscores the need for this reflective scaffolding. In the unprompted case, people frequently default to optimistic scenarios, overlooking mitigation strategies for content they would not enjoy. The planning fallacy highlights how people systematically overestimate their ability to foresee complex futures, producing confident but incomplete plans that particularly overlook problems~\cite{kahneman1977intuitive}. The illusion of explanatory depth helps explain this overconfidence in planning: people often believe they understand mechanisms or preferences in rich detail, but when pressed to explain or operationalize them, their understanding collapses into vagueness or contradiction~\cite{rozenblit2002illusion}. In the feed specification case, users may confidently articulate what they want (“funny posts,” “informative news”), but have not followed those instructions to their logical end to realize the flaws in the resulting feeds. Without being prompted to reflect and revise, their specifications risk being overconfident and underspecified.

Information science has similarly had to contend with this struggle of user specification elicitation in the domain of information-seeking, and arrives at a similar solution to ours. The four levels of information need articulation describe that an information need develops through successive phases: it begins as an internal sense that something is missing and only later becomes something that can be stated in explicit terms~\cite{Taylor1968QuestionNegotiationAI}. Early formulations are fuzzy and unaligned with the true need because the underlying idea has not yet been organized enough to express directly. Later stages distinguish the clearest expression of need from the eventual request, which has been reformatted to anticipate what is available to request. The anomalous states of knowledge theory explains the same problem in terms of knowledge structure: when someone tries to specify what they want, they are doing so while their understanding is still incomplete, unstable, or internally misaligned, so the statement cannot capture the actual need~\cite{belkin1982ask}. In our domain, users describing their feeds will often first produce a surface-level expression shaped by conversational pressures, available vocabulary, and their understanding of their needs and wants. They may say they want ``funny posts'' or ``important news,'' not because those phrases capture a settled and clearly articulated preference, but because those are the most readily available way to convey something they cannot fully articulate yet. It takes iteration grounded in examples to understand the insufficiencies of that earlier communication. Users need to grapple with what they want at a more specific level, especially in terms of the available content. However, in the feed context, unlike in information seeking, users are not necessarily addressing true ``needs'' with imagined utility beyond entertainment where success can be readily evaluated, so they are even more likely to diverge from their earlier intentions through the iteration process. Similarly, the interview-based iteration may be more of an exploration or a pleasurable reflection, in and of itself.

The “reference interview’’---the structured but conversational process through which a librarian uncovers a patron’s true information need---is their response to this specification issue. These interviews explicitly treat initial user statements as provisional, incomplete, and often misleading; thus, effective interviewing requires iterative probing, clarification, and reformulation to surface latent goals and constraints~\cite{katz1969introduction}. Later theorists have framed the reference interview as a collaborative meaning-making activity in which both librarian and patron jointly refine the problem definition, with the librarian interviewer responsible for eliciting tacit preferences, exposing ambiguities, and helping patrons recognize unstated assumptions~\cite{brown2008reference}. Like our feed preference elicitation setting, reference interviewing acknowledges that preferences and needs are often discoverable only through interaction, and that carefully scaffolded dialogue can compensate for cognitive gaps that make initial self-specifications unreliable.

\section{Feed Elicitation}

Our principal aim is to elicit users' desires for their social media feeds so that we may create personalized feeds that better suit their needs and preferences. We therefore require not just a feed elicitation process, but an accompanying feed creation algorithm both to evaluate the success of this elicitation and to realize its utility.

We separate out these two aims into distinct steps: an elicitation interview, and a two-part post classification and rating algorithm. These steps interface with a natural language feed description, which serves as the output of the elicitation interview and the input to the post-classification system. Both of these steps are currently enabled by LLMs, but the classification step could later be carried out by a smaller, distilled model for more scalable deployment. Below, we describe our design goals as well as the practical implementation of this system.

\subsection{Elicitation Aims}

Our elicitation process seeks to uncover what content people actually want to see. We aim to help people thoughtfully consider what their actual ideal feed would be, rather than fixating on the functionality already familiar and available (e.g., topic-based sorting) and leaving out details critical to their satisfaction. 
We believe people struggle to fully and clearly detail their desires and end up underspecifying what type of content they would prefer. Users expect behavior like existing social media platforms: topic- and keyword-based search, pre-preemptive filtering of low-quality or highly toxic content, and little else in the way of quality control except for engagement signals. Their specifications then reflect a fixation on feeds that map into that paradigm.
Users' fixation manifests in several ways: a predisposition towards broad topical specifications, missing removal criteria for content that would be disruptive or distressing, and a lack of attention to what qualities determine enjoyment or benefit. 

We design this elicitation process to force users to actively consider their preferences for key qualities that are potentially relevant to their feed context. 
Users are directed towards considerations beyond topic and engagement metrics, like the purpose that posts in the feed should serve and the way they want to feel while using the feed. The interview prompts decisions around additional content to include or exclude, and uses frequent and plentiful contextualizing examples to guide exploration of the design space under consideration. It therefore aims to (a) provoke consideration of factors beyond topic, (b) alert users to possible edge cases for their feed, and (c) ground these preference details in the domain and possible content.

The feed elicitation uses an interview in order to be maximally adaptive. In order to appropriately ground user specification, the questions are accompanied by examples and suggestions tailored to the users' current specification and its domain(s). Rather than a static set of questions or instructions, the system can give contextual suggestions and clarify when specifications are unclear. 

We choose natural language as an input domain for feed specifications because it maximized flexibility with respect to controlling the actual implementation, which is currently LLM-based. While we aim for a future with less costly and computationally-heavy implementations, LLM classification is the technology that currently enables this level of expressivity in feed creation. 
While we could allow interaction at an abstracted or concept-based level, that interaction would be translating into natural language at some point, and therefore denying the user a level of closer and more fine-grained control. Existing implementations may operate with block-based and pre-set controls~\cite{graze2025, skyfeed2025}, but those alternate inputs restrict users to only being able to express in terms of the functionality provided. Our goal is to take advantage of the greater expressivity in feed control enabled by LLMs, and help users make use of its full range.

\subsubsection{Interview}

\begin{figure*}[tb]
  \centering
    \includegraphics[width=0.9\textwidth]{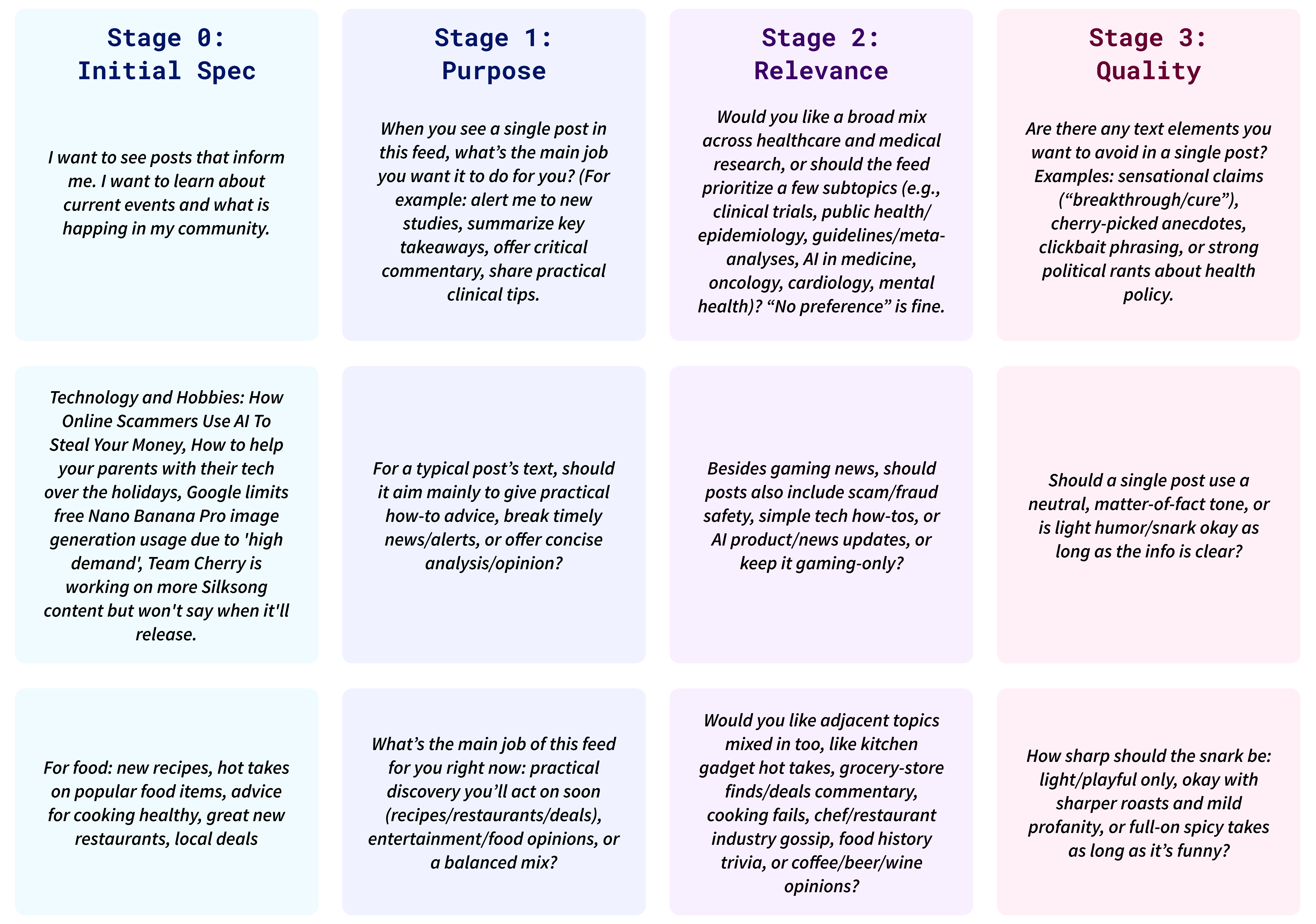}
  \caption{The elicitation interview proceeds in three main stages, meant to direct user intention towards the different main areas of interest for their specification}
  \label{fig: interview process}
\end{figure*}

We divide the interview into three main stages, designed to query the purpose of the feed, the main topics of relevance, and then desirable qualities for posts in the feed. Each stage is defined with a set of informational goals which should be ascertained during that stage. Example questions from each stage are shown in Figure~\ref{fig: interview process}.
The interviewer is prompted at the beginning of each stage with those goals, as well as a set of interviewing principles to ensure a useful rather than aggravating interviewing experience (which stays constant across stages).

During the interaction, the interviewing system poses one question at a time to elicit information relevant to the above stage-specific goals. After each question response, users must also rate how important it is for the feed to follow the preference they just expressed (Mildly Preferred - Preferred - Essential). Then, the system applies a reflection prompt to assess whether the accumulated dialogue is sufficient to address all goals for the current stage. If so, the system proceeds to the next stage. 

The initial, purpose-focused stage, is designed to set the stage for the interview and prime the user to be thinking about how the feed could best serve them: what do they want to accomplish, and what would be advantageous or disadvantageous towards that goal. Purpose is goal-oriented, focusing on a higher level of abstraction than topic or other content-based attributes, trying to ascertain what they are hoping to get out of the posts in their feed. For example, this stage might uncover that the user wants to learn interesting facts that leave them excited about the world or that they want to feel soothed by relaxing content. Following this stage, the interviewer has information about the user intentions to help it predict their potential preferences to contextualize the coming questions and offer better examples. 

The next stage, focused on topic, aims to establish what content is relevant. The questions prod the users' decision barrier of what they do and do not want to see in their feed. Specifically, this stage is supposed to uncover if there are edge cases of desired topics that they have not thought of, or of undesired topics that are currently considered in-scope. For example, this stage might find that a user is interested in learning about science, especially astronomy, but not human biology or things that gross them out.
Finally, the third main stage focuses on other post attributes---framing, emotional impact, etc.---that users may have preferences on. It aims to prompt the user to consider if these different factors matter for their satisfaction or enjoyment. For example, this stage may uncover that the user prefers fun facts over recent scientific discoveries and that they want to prioritize accessibility and novelty rather than depth. 

\begin{figure*}
  \centering
    \includegraphics[width=\textwidth]{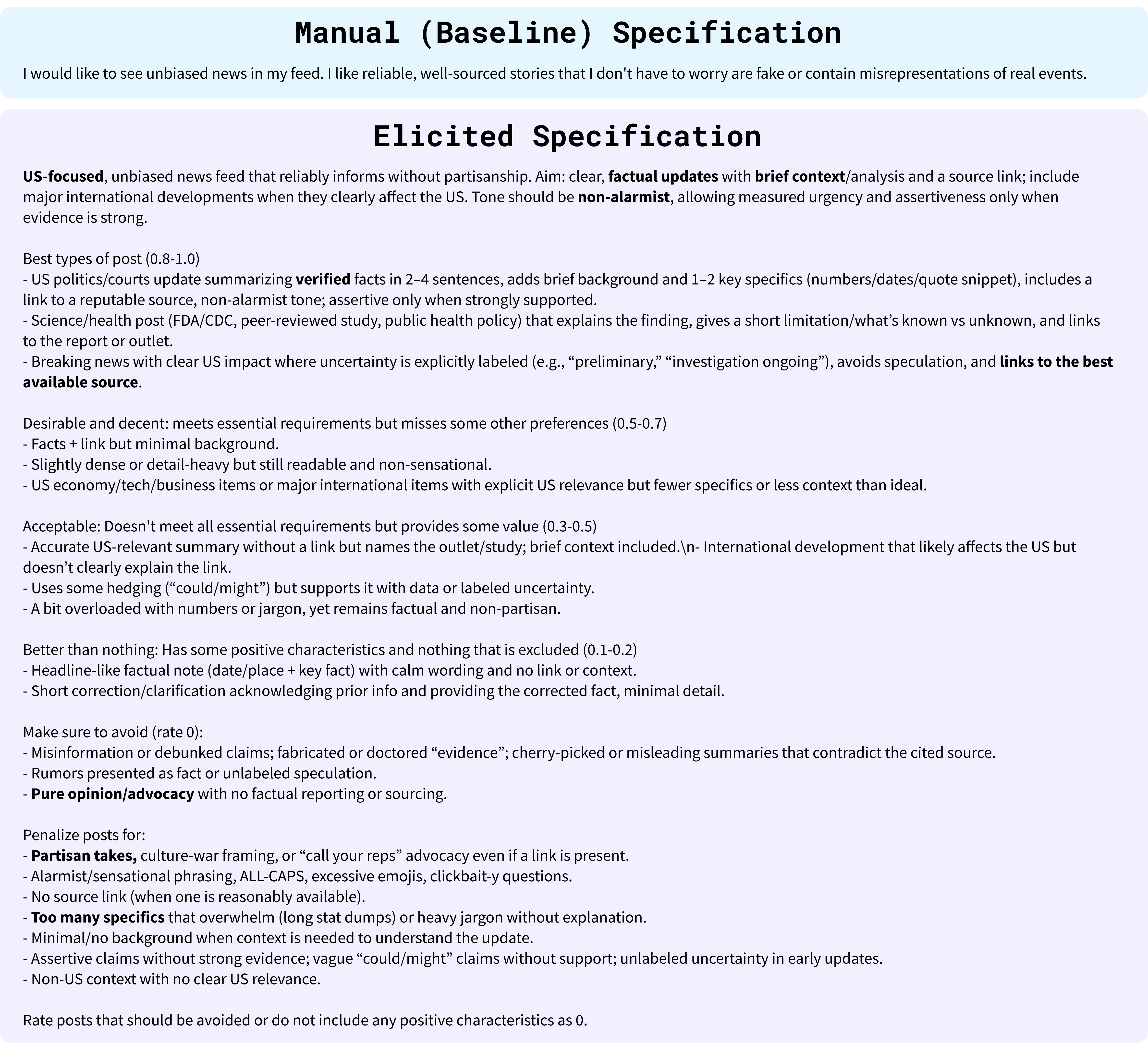}
  \caption{The feed elicitation process produces thorough specifications with clarifying details about the feed intention, preferred quality attributes, ideal types of content, and content to remove or downrank. (Bolding added by researchers for emphasis of notable additions.)}
  \label{fig: specs}
\end{figure*}

Following, the interview allows the user to provide additional information or corrections, before outputting a final specification from the conversation, though further corrections are allowed by the system. The final specification considers the entire conversational history to synthesize a final structured feed specification. Prompts for all steps of the interview process are included in Appendix~\ref{app: prompts}. We give an example of a paired initial specification and interview output specification in Figure~\ref{fig: specs}

\subsection{Feed Creation}

Next, we need to transform these specifications into a systematic classification and ranking algorithm.

Drawing on practices from information retrieval~\cite{manning2008introduction}, we separate out \textit{relevance} from \textit{ranking}. So, from the interview data, we construct two prompts, one designed to serve as a first-pass filter of post relevance, and the other to rank content's utility towards the users goals for the feed (a ``quality'' rating). If these two phases are combined together into a single prompt, we observed that content that violated specific topic preferences would be rated highly for fulfilling other desired criteria, when the specific articulated preference logic required that said content always be excluded. We therefore perform an initial broad relevance pass that rules out specifically detrimental or completely irrelevant content, and then a secondary quality rating that includes the more nuanced preference and value calculations. 

We therefore create two classification schemes off of the preference data. The initial post relevance classifier is designed to filter very broadly for relevance by expanding the subjects found in the existing specification to a broader set of even potentially relevant topic areas. The quality ranker uses the feed specification faithfully but appends rating instructions. Further details of how the prompts are used are found in Appendix~\ref{app: classifier prompts}.

In order to successfully filter and rank posts, the relevance classifier is run first, and then the quality ranking is run only on posts that are classified as sufficiently relevant. No additional classifying logic (e.g., engagement, author relationship, recency) is currently included; the quality ranking is used alone, though additional logics could be introduced given the desire. Currently, this system operates on a static database, though it could be adapted to perform classification on incoming posts through the BlueSky Firehose, or on users' existing feed posts on other platforms.

\section{Evaluation Method}

We propose \textit{feed elicitation interviews} with the goal of guiding users to create feeds of content they actually want to see, and that support them in their various aims for social media use. Our main research questions are as follows:

\textbf{RQ1:} Do users prefer the feeds produced by the elicitation interviews over those manually specified?

\textbf{RQ2:} How do users experience the elicitation process in terms of cognitive effort, enjoyment, and satisfaction?

We therefore want to show that this feed elicitation method produces feeds that accurately reflect users' preferences, and that are preferred to baseline approaches such as prompting. We address this question of preference at a feed level (which feed do they prefer?) as well as on a granular, post-level basis (do they like each post). We also want to show that the elicitation interview does not present extreme additional effort or cause frustration for users. 

We ran an initial study on a previous iteration of our system with some slight methodological iterations. Full details of the procedure and results are included in Appendix~\ref{app: old study}.

\subsection{Participants}
We recruited $N=400$ participants from Prolific (\url{prolific.co}) and Cloud Research Connect (\url{https://connect.cloudresearch.com}), two online crowdsourcing platforms. We determined our sample size through simulation-based power analyses using effect sizes estimated from preliminary data. Because our design includes post-level and feed-level comparisons, we simulated power for each planned analysis separately. To achieve our target of 90\% power, the post-level approval comparisons required approximately $N=120$ participants per condition, while the feed-level preference comparison required approximately $N=200$ participants per condition. We therefore set our recruitment target to 200 participants per condition.
We selected participants that had at least 50 submissions, were located in the United States, were native English speakers, had an approval rating of at least 95\%, had not completed our task before, and had some regular use of Twitter/X (the only text-and-feed-based social media available as a filtering option). All participants received a payment of \$11 for approximately 40 minutes of their time, resulting in pay at slightly over \$15 per hour.

\subsection{Procedure}

\begin{figure*}[tb]
  \centering
    \includegraphics[width=0.9\textwidth]{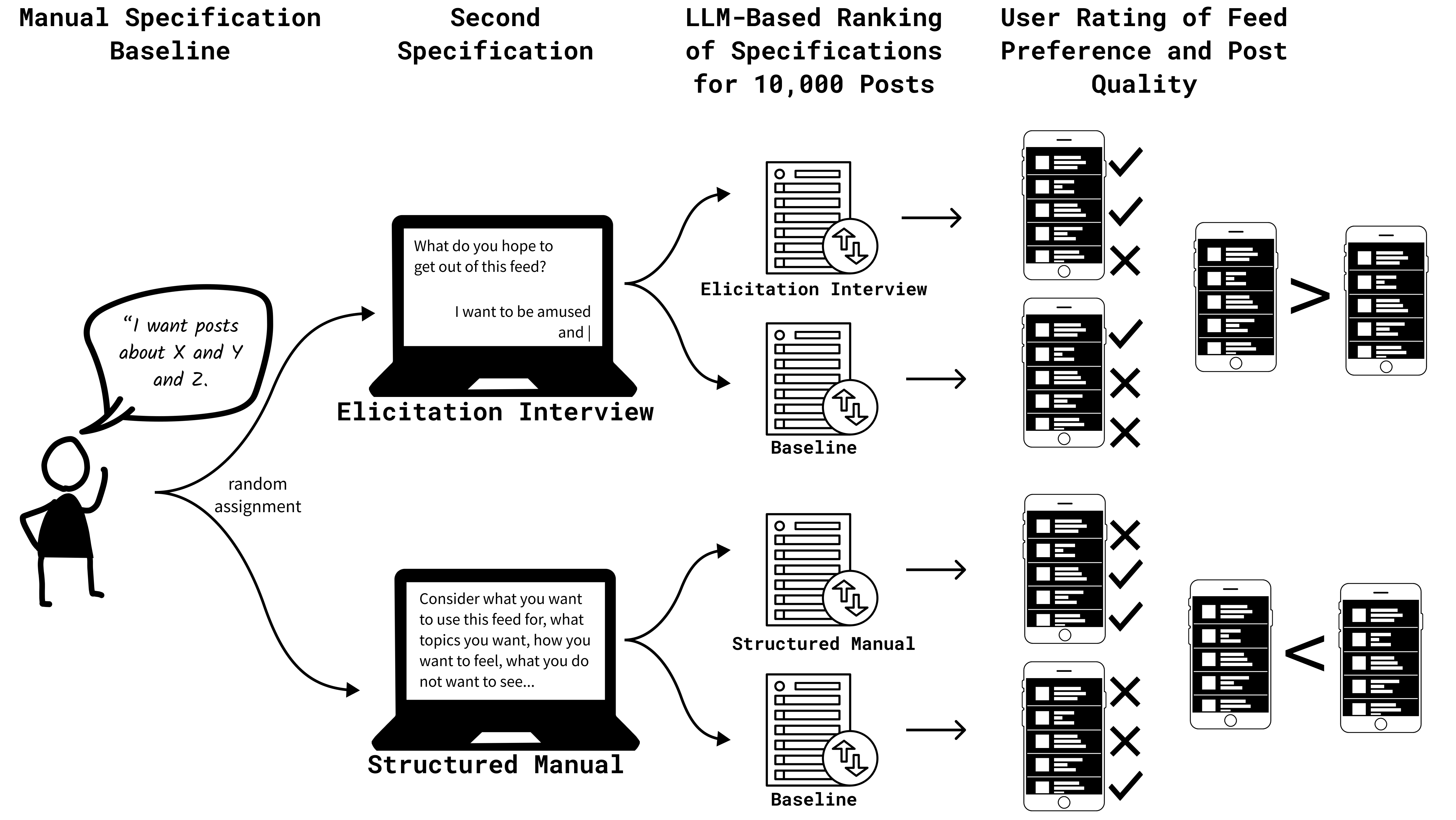}
  \caption{Each participant completes two different types of feed specifying activities, one being a baseline. Then they perform various quality rating and ranking tasks to evaluate overall quality and preference.}
  \label{fig: evaluation process}
\end{figure*}

All participants start out on a page explaining the idea of feed creation. They are told that they will be constructing two versions of a feed designed to support them in some area of their social media use. They are first prompted to imagine a few (3-5) different genres of feeds that they might enjoy, were they to divide their browsing into different categories. 

Starting from one of those feed types, their next task is to complete the baseline feed specification task, describing what types of posts they would like to see in that feed. Then, they are randomly assigned to one of two enhanced elicitation methods, which they complete to create a second feed specification. Two different feeds---the baseline, and the treatment condition that they are assigned to---are then generated according to the participant's preferences, which are randomly ordered and then presented together, side-by-side, to them. During the generation time, we ask the users to reflect on their feed description experience, both in a free-response question and through the NASA-TLX measures for mental demand, effort, frustration level, and success.

When comparing the feeds, participants are first asked Likert-style questions about each of the twenty top posts in each feed (a total of forty posts). They are asked to rate their approval of that post on a 7-point scale (Strongly Dislike - Neutral - Strongly Like). After seeing both feeds, they are asked overall preference between the two on a seven-point Likert scale (Strongly Prefer Feed 1 - Prefer Feed 1 - Slightly Prefer Feed 1 - Neutral \ldots) and to explain their preference. 

\subsubsection{Conditions}

There are three different feed-elicitation methods that participants may undergo. We have one baseline condition (where the user manually specifies their feed preferences with minimal instruction) that each user completes at the beginning. Then, they are randomized into one of two improved feed elicitation methods. The experiment is a between-subjects study, where each participant produces two versions of a feed: one using the baseline, and one using their assigned treatment condition:
\begin{itemize}
    \item Manual (baseline control): After listing a few plausible types of feed, users are told to ``Now, choose ONE of the categories you listed earlier to design a feed for. Consider that we cannot accommodate extremely niche topics, posts from specific people (like those you know), or image/video-based content. Below, describe the type of posts you want to see in your feed. You will not be able to proceed without writing something.''
    \item Structured manual (stronger control): Users are straightforwardly given similar goals as the interviewer agent: ``You are describing a custom feed based on the feed topic you selected earlier. Reminder that we cannot accommodate extremely niche topics, posts from specific people (like those you know), or image/video-based content. Consider what you want to use this feed for, what topics you want to see, how you want to feel using it, and any other characteristics that seem relevant. Also consider content you want to prioritize or avoid. Below, describe the type of posts you want to see in your feed. You will not be able to proceed without writing something.''
    \item Elicitation interview: Users are questioned by an interviewer agent that proceeds in three stages to query them about their purpose for this feed, their desired topics, and their preferred framing (e.g., emotional, factual, etc) of the posts. These preferences are synthesized into a final specification. 
\end{itemize}

The structured manual description condition serves as an ablation of our full interview procedure, removing the interactive nature of an interviewing bot with simple instructions which convey the logic of the interview procedure. We believe that the nature of an interview itself, rather than the emphasis on describing multiple aspects of desired content, results in the users reflecting on their wants and describing them more specifically. However, we use this condition as a strong control, to measure the gains of an interactive rather than static process of elicitation.

We employ this mixed study design to best isolate the impact of the feed elicitation without being able to change order of conditions. 
We maintain a point of reference for the feed improvement via the baseline condition while mitigating some of the uncertainty around order effects.

\subsubsection{Evaluation-Specific Implementation}

For the purpose of the evaluation we conduct, we generate a feed out of a static database of 20,000 posts downloaded from the BlueSky Firehose API~\cite{blueskyfirehose2025}. These posts have been pre-filtered to remove NSFW posts (classification performed with a zero-shot prompt, see Appendix~\ref{app: nsfw prompt}), those with only hashtags or links, and those with fewer than three words. We run the relevance classifier on 10,000 posts initially, and run it on an additional 10,000 if fewer than 100 are found to be potentially relevant. This process takes fewer than five minutes for each feed, and is performed in the background for the baseline feed (while the user completes the other elicitation procedure), so participants are only made to wait for processing once. 

\subsection{Analysis}

Through our analysis, we aim to test whether feed elicitation selects content that users prefer to content in other feeds.
We also want to show that this content difference produces a difference in feed experience that registers to users. 
Our preference data comes in two forms: overall feed preference (do they prefer the baseline or their assigned condition feed) and approval of each post in each feed. We analyze this data to address \textbf{RQ1} with feed- and post-level preference. Then we analyze the free response text from users to better understand their feed description experience (addressing \textbf{RQ2}), as well as what factors contributed to their satisfaction or lack thereof with the feeds produced.

\subsubsection{Feed-level Preference}

To analyze feed-level preference, we perform simple non-parametric tests, using a Wilcoxon signed-rank test for the within-subjects comparison of interview-based feed versus baseline and a Wilcoxon rank-sum test (Mann-Whitney U) to analyze between-subjects difference in preference---comparing the magnitude of preference between the two treatment conditions (elicitation interview versus structured manual) over the baseline. 

\subsubsection{Post Quality}

As an initial descriptive comparison of post-level ratings, we conduct paired Wilcoxon signed-rank tests within each participant group to assess whether the experimental feed differed from its corresponding baseline. This provides a straightforward indication of whether the rating distributions shift under each condition. Because participants rated only one experimental feed, direct comparisons between the two experimental feeds are not statistically identifiable.

While Wilcoxon signed-rank tests provide a nonparametric within-participant comparison of each experimental feed to its corresponding baseline, they do not allow us to model post-level ratings directly, incorporate participant-level dependencies, or estimate effects of both feed type and participant condition. To account for the ordinal structure of the ratings, the hierarchical nature of the data, and the need to estimate predictor effects on the full distribution of post-level evaluations, we fit a cumulative logit mixed-effects model. This approach allows us to model individual posts as the unit of analysis while accounting for repeated ratings nested within participants.

To assess whether a cumulative link mixed model is appropriate, we first must assess whether the proportional-odds assumption holds. We therefore apply a scale test to a corresponding fixed-effects cumulative link model using the same fixed-effects structure as the primary model, which evaluates the proportionality of the predictors’ effects across cumulative thresholds. 
Then, we fit a cumulative logit mixed-effects model with feed type (baseline, structured manual, elicitation interview) and participant condition (structured manual or elicitation interview), along with a random intercept for participant to account for participant-level rating tendencies:
\begin{verbatim}
    approval ~ feed_type + participant_condition 
               + (1 | participant_id)
\end{verbatim}
Because no participant rated both experimental feeds, direct contrasts between the structured manual and elicitation interview conditions are not identifiable and are therefore not estimated.

\subsubsection{User Experience and Interview Process}

We ask all participants to reflect on their feed description experience, both through an open-response question and through the NASA-TLX measures (excluding physical demand). To compare experiences across structured manual and elicitation interview experiences, we perform unequal-variances t-tests with Holm correction. We then inductively code their free response reflections to understand the general patterns of frustrations or struggles.

We also look to the users' written feedback about their experience and the feeds they saw to contextualize their preference. We perform inductive coding on the different explanations of user preference to understand the major reasons users preferred one feed over another. We follow by lightly coding the elicitation interviews to understand possible failure or success cases of the interview. We look for signs of user frustration or satisfaction, and then try to understand the patterns of behavior that produced these reactions. We also look for questions that prompted responses that produced key outcomes in the eventual specification.

We then analyze themes in the intermediate artifacts produced by our participants: the feed specifications. We look for characteristics either common in feed specifications or that we aimed to produce via the specification process: general desire for feed experience, topic(s) of focus, topics to avoid, content characteristics to avoid, characteristics of desired content. Here, we aim to see if the interview had the desired impact in prompting users to consider and articulate their feed preferences comprehensively.

\section{Results}

\subsection{Participants preferred feeds produced by elicitation interviews}

\begin{figure*}[tb]
  \centering
    \includegraphics[width=0.8\textwidth]{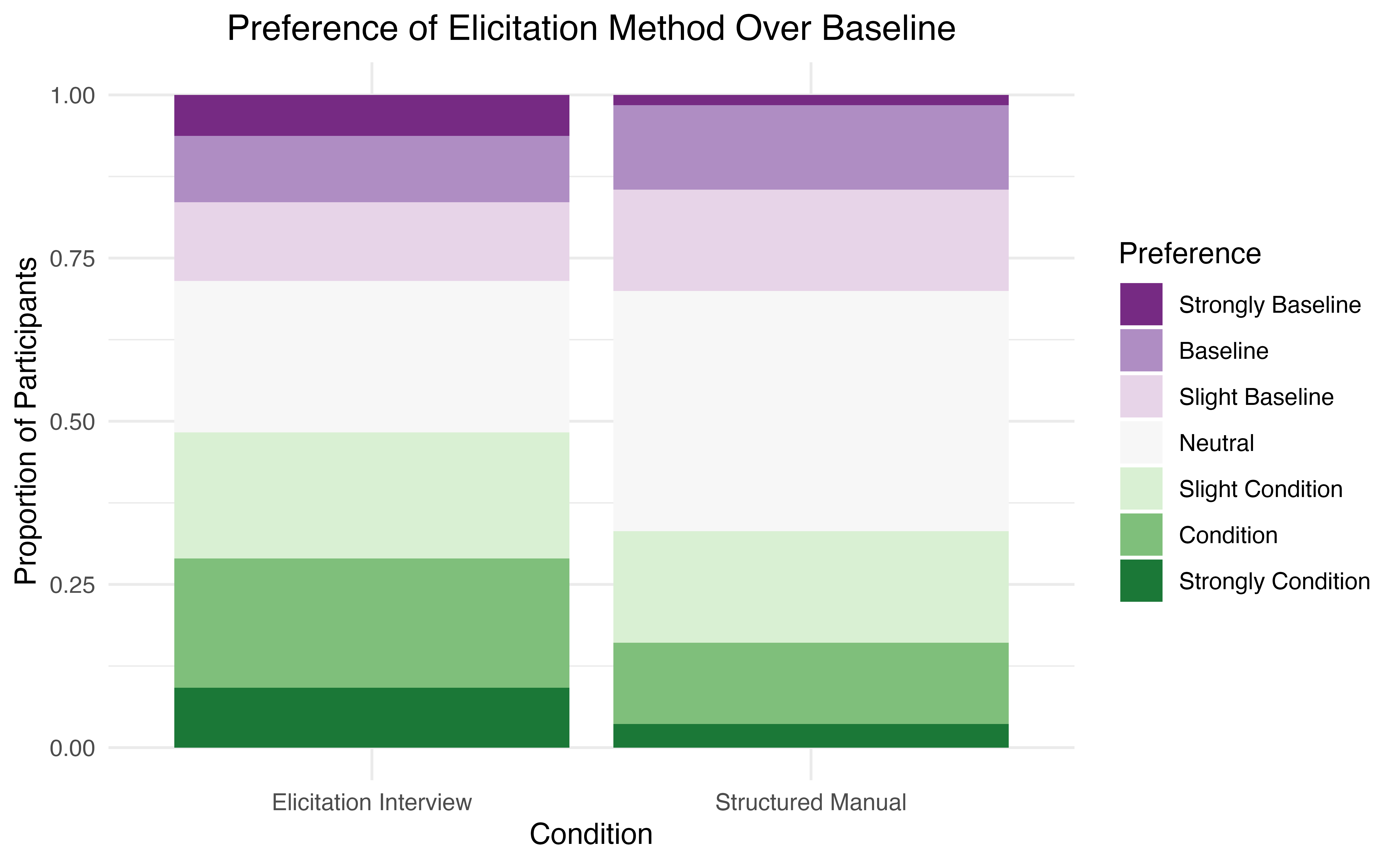}
  \caption{Participants overall prefer the elicitation interview over the baseline and show no significant preference either way for the structured manual specification versus the baseline. However, participants who completed the interview lean towards more extreme feelings in both directions, feeling strong preferences for and against the interview condition at higher rates. Over a third of participants had no preference between the two manual specifications, whereas closer to a quarter had no preference between the interview and baseline feeds.}
  \label{fig: new feed preference}
\end{figure*}

Our participants preferred the interview-generated feeds 48\% of the time, more frequently than they held neutral preference (23\%) or preferred the baseline (29\%). While the preference perspectives were relatively polarized, with a notable amount also preferring the other feed, we see an overall preference toward the feeds produced by the elicitation interview. 
In comparison, participants who completed the structured manual condition and baseline were almost evenly split between preferring structured manual (33\%), no preference (37\%), and preferring the baseline (28\%). The full distribution of preference rating is visualized in Figure~\ref{fig: new feed preference}.

To test whether these preference patterns departed from neutrality, we conduct paired Wilcoxon signed-rank tests within each participant group. For the elicitation interview condition, preference scores were significantly greater than zero ($V = 8003$, $p = 0.004$), indicating a reliable overall preference for the interview-generated feed over the baseline. In contrast, preference scores in the structured manual condition did not differ significantly from neutral ($V = 3990.5$, $p = 0.53$), consistent with the more evenly distributed preferences observed descriptively. We also compare preference scores between the two participant groups. A Wilcoxon rank-sum test reveals a modest but significant difference between the elicitation interview and structured manual conditions ($W = 22462$, $p = 0.028$), indicating that participants who completed the elicitation interview expressed stronger preference away from the baseline than those who completed the structured manual specification. 

We turn to analysis of post-level rating data to see if the feed-level preference is consistently mirrored in ratings of individual pieces of feed content.

\subsection{Participants liked the posts more in feeds resulting from the elicitation interviews}

\begin{figure*}[tb]
  \centering
    \includegraphics[width=0.8\textwidth]{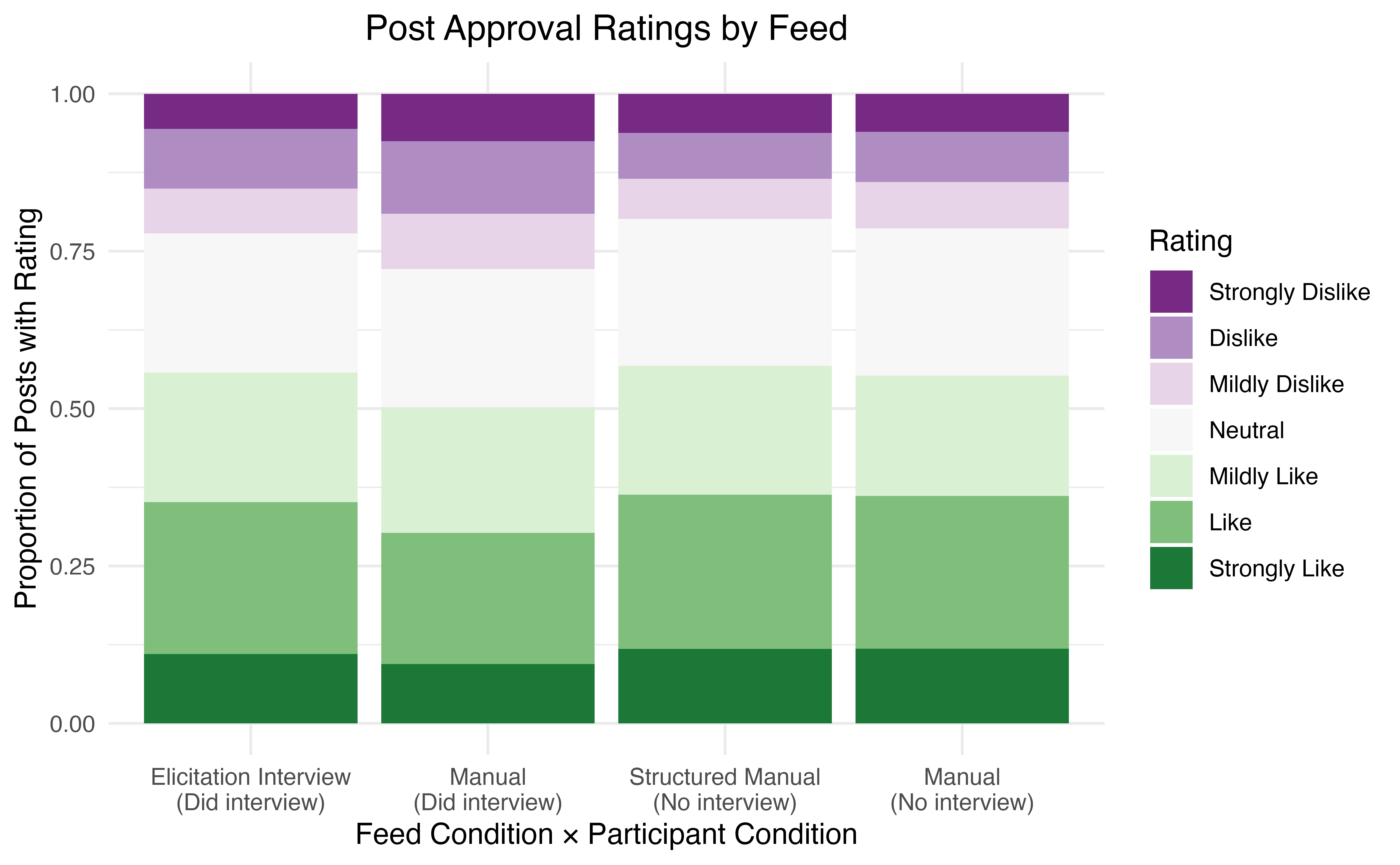}
  \caption{Participants approved the posts at a modest but significantly higher rate for feeds created by the elicitation interview as compared to the manual specification. }
  \label{fig: new post rating plot}
\end{figure*}

Participants rated posts from the elicitation interviewer feeds higher than the baseline at all levels of the scale. 
Though all feed conditions saw more posts approved than disapproved, the elicitation interview had a higher proportion of approved posts (56\% versus 50\% for baseline) and fewer disapproved posts (22\% versus 28\% for the baseline). This corresponds to a reduction by about a quarter of disliked content and an increase of over 10\% of approved content versus the baseline. Exact breakdowns of the approval rates are shown in Figure~\ref{fig: new post rating plot}.

To formally assess whether these apparent differences in post-level ratings differed across condition, we conduct paired Wilcoxon signed-rank tests. For the elicitation interview condition, the median shift in post ratings was positive, indicating that participants generally evaluated posts more favorably after completing the interview. This improvement was statistically significant ($V = 11798$, $p = 0.0017$), confirming that the upward shift in ratings reflects a consistent pattern across participants rather than sampling variation.

For the structured manual condition, however, post-level ratings did not differ systematically from baseline. The median change in ratings was neutral, and the Wilcoxon signed-rank test (with continuity correction) does not reveal evidence of a departure from baseline evaluations ($V = 8718.5$, $p = 0.49$). Thus, while the elicitation interview led participants to rate individual posts more positively, the structured manual condition did not produce a comparable shift.

However, notably, the ratings were visibly higher across all feeds---both experimental and baseline---for participants who never completed an interview. We therefore look to a richer model to help explain this result. Since participants rated both the baseline feed and the treatment feed at the same time after authoring their feeds, the ratings on the baseline posts provide a point of overlap that allows us to assess whether the interview itself alters participants' use of the rating scale. A cumulative link mixed model including participant condition as a predictor shows a significant shift in baseline rating behavior between the two groups: participants assigned to the structured manual condition gave higher ratings to the baseline posts than did participants who completed the elicitation interview ($\beta = 0.31$, $p = 0.012$). The odds ratio indicates that participants assigned to the structured manual condition were approximately \emph{1.36 times} as likely to give baseline posts a higher rating than participants who completed the elicitation interview. In other words, participants who experienced an interview gave lower ratings to \emph{all} feeds versus those who experienced the control condition, perhaps because the interview prompted deeper reflection on what participants actually wished to see. 

However, controlling for baseline scores, the model confirms that the interview condition produced the highest-rated feeds. The same model shows that, relative to the baseline feed, posts generated from the elicitation interview were rated significantly more positively ($\beta = 0.25$, $p < 0.001$), whereas posts produced using the structured manual specification did not differ reliably from baseline ($\beta = 0.06$, $p = 0.17$). Because participants rated only the treatment feed associated with their assigned condition, these effects reflect within-group shifts relative to each participant’s own baseline rather than direct contrasts between the two experimental feeds. This model indicates that while the elicitation interview led to more conservative rating behavior, it also produced posts that received higher latent approval (where posts were 1.28 times as likely to be rated higher) than the baseline, whereas the structured manual specification did not produce a detectable shift in post-level approval relative to baseline. 

To ensure that the cumulative logit specification was appropriate, we evaluate the proportional-odds assumption for the fixed-effects structure. A scale test shows no evidence of nonparallel slopes for either predictor (both $p > .28$), supporting the use of our selected model.

From this analysis, we can tell that the interview helped generate feed specifications that more closely reflected participant content preferences. Participants preferred content produced by the elicitation interview to that from the baseline, while structured manual specification demonstrated no similar significant trend. Qualitative feedback reflected improvements in post preference as well. Participants praised the elicitation interview feed for being ``exactly what I was looking for'' with ``all the things that I wanted to see and read.'' ``exactly spot on,'' and ``almost exactly what I wanted,'' in contrast to the baseline alternative, ``not what I asked for at all.'' However, at the same time, completing the interview seems to have made participants more exacting in their taste, likely because they were forced to consider and articulate exactly what they want and do not want out of their feeds.

\subsection{Participants appreciated additional specificity and personalization in feeds produced via interview elicitation}

When participants had distinct preferences that they failed to communicate, they really appreciated the difference in their elicited feeds. One participant described the difference:
\begin{quote}
    Feed 2 [the feed produced by an elicitation interview] contains more posts that are actionable and practical, such as the Pomodoro tip and study routine post. It aligns better with my core interests in habits, focus, and productivity, and avoids vague motivational or overly reflective content that doesn’t suggest a clear action. Feed 1 [the baseline feed] contains several posts that are primarily reflections or vague motivational statements, which are less useful to me.
\end{quote}

Other participants remarked on how much the interview personalized to them, noticing that the feeds were ``more of what I had in mind for my feed'' with ``some politics, science, etc. just as I had requested,'' or had ``more topics that i relate to [sic].'' In these cases, the baseline feed sometimes served them content that felt like obvious mistakes, like having ``empty fluff posts,'' ``off-topic content, such as economics papers and general news,'' or ``irrelevant weather alert.'' Participants frequently left critical requirements unstated in their initial specifications, like stating that they wanted ``very funny videos that have high quality editing and jokes'' and then preferring a feed that served ``more political satire tweets'' versus ``comedic posts that weren't related much to current news and politics.'' The interview feed helped specify many of these types of preferences, like wanting food content to ``help audience to understand more about food [sic],'' media takes that were ``concise and very clear,'' and humorous posts with ``snappy one line jokes and clever wordplay.''

The interview also helped participants remember to omit relevant but disruptive content that they appreciated not seeing in their interview-produced feeds, like ``flag burning,'' ``self-promotions,'' ``a dog being sick,'' and ``posts that were just advertising links.'' The full process of making these specification adjustments is shown in Table~\ref{tab:interview-improvements}, which starts with the initial feed description, and shows how a particular interview question resulted in a positive change for the feed.
These observations align with the quantitative rating data, where the elicitation interview produced more desired posts and fewer disliked ones, both tailoring to their interests and removing the content that they had not thought to rule out.

 \newcolumntype{L}[1]{>{\raggedright\arraybackslash}p{#1}}
 
\begin{table*}
  \centering
  \renewcommand{\arraystretch}{1.6}
  \small
  \begin{tabular}{L{1.3in} L{1.1in} L{1.2in} L{1.1in} L{0.9in}}
  \hline
    Initial Specification & Question & Response/Preference & Specification & User Reflection \\
    \hline
   I want to see real-time game updates, scores and highlights. I am interested in injury reports, trade rumors, signings, and analysis from sports journalists. I would also like to see interesting stats and historical facts regarding different sports.
    & Which sports or leagues do you want real-time updates for? You can list a few priorities or say “any major sports”—no need to be specific if you don’t want to be.
    & NFL, NBA, MLB, UFC. (preferred)
    & A real-time, multi-league sports feed that prioritizes breaking updates for NFL, NBA, MLB, and UFC.  \ldots Avoid entirely:\ldots Posts not about sports or not about NFL, NBA, MLB, or UFC (unless clearly tied back to those leagues). \ldots
    & I just liked the sports talked about way much more. I did not like the first feed [baseline] at all as most of it was about soccer.
    \\
    I want to see posts that inform me. I want to learn about current events and what is happing in my community.
    & What city or region should “my community” refer to in the posts?
    & the northeast ohio area. (preferred)
    & A neutral, calmly-informing feed of Northeast Ohio current events. Prioritize useful local updates on weather, economy, new businesses/development and road work, and community events. Include Ohio/national items when the local impact on Northeast Ohio is clear.
    & Feed 1[interview feed] seemed to have many more items that were relevant to me, and less stuff related to politics. I liked the local focus of 1[treatment feed], while 2[baseline] had more irrelevant weather alerts.\\
    
    Sports. NFL-Buffalo Bills. MLB-Seattle Mariners
    & For the tone of quick updates, do you prefer strictly neutral/factual wording, or is some fan excitement/hype okay? (You can say no preference.)
    & factual. (essential)
    & A fast, factual update feed focused on the Buffalo Bills (NFL) and Seattle Mariners (MLB) \ldots  Make sure to avoid \ldots opinion-only/hype/banter with no clear factual update \ldots
    & Feed 2 [treatment feed] seemed to have more exact information whereas feed 1[baseline] was a lot of opinions\\
    &&&& \\

    \hline
    \\[-1.8ex]
    \end{tabular}
    \caption{The interview gives users a chance to clarify underspecified points from their original specification. These clarifications had significant impacts in the resulting feeds and produced qualitatively better browsing experiences for the participants.}
    \label{tab:interview-improvements}
\end{table*}

Specificity was not a purely beneficial characteristic, however. 
Multiple participants critiqued the interview-produced feed for overly focusing to the point of reducing content diversity. A sizable proportion of participants who prefer the baseline feed do so for this reason: the elicitation interview feed is ``extremely repetitive,'' while the baseline feed is ``a bit more balanced'' and ``broader,'' with ``lots of different posts'' and ``more of the things i liked [sic].''

As implemented, our elicitation interview's tendency towards overspecification may rule so many posts as to leave only low-quality content. A common complaint against the elicitation interview feed was that some of the posts were irrelevant and not what they wanted. As one participant experienced, ``[The baseline feed] had more Star Wars related content. [The interview feed] had some that didn't seem related at all,'' even though they had expressed a desire for Star Wars content in both. In these cases, due to a lack of inventory focused on the most important preference (in this case, relating to Star Wars) means that additional expressed preferences may dilute the signal of that initial preference and therefore produce a less desirable feed. While the specification holds some of the blame, the available inventory of posts shares responsibility. We set significant mitigation steps for overspecification into the interview: direct instructions to accept broad user preferences, a preference-rating step after answers that requires participants to articulate how much they care about each preference being followed, limits on the number of questions asked, and a relevance rating step that deliberately broadens the scope to identify potential posts. However, the interview is not grounded in the available inventory of posts and may therefore promise a user the ability to select incredibly niche nonexistent posts that match their wants exactly, only for that specification to rule out many posts they would have been satisfied with.
An improved system could either search through a larger set of posts or give feedback to the user when it failed to retrieve significant quality content.

The elicitation interview can benefit people who have left out critical details of what they want from their feed. However, the appropriate specificity depends significantly on the user's preferences, so it is difficult to correctly control.

\subsection{Interview process took additional time but not cognitive effort}

Participants in the structured manual and elicitation interview conditions reported broadly similar levels of overall workload, but several NASA–TLX subscales showed clear differences between conditions. Mental demand and effort did not differ reliably, suggesting comparable cognitive requirements across the two workflows. However, participants in the structured manual condition reported higher temporal demand and frustration than those in the elicitation interview condition, though both were very low for each condition: mean temporal demand was $3.9$ versus $2.8$ on a scale from $0-20$ and mean frustration was $4.0$ versus $3.4$. Performance ratings showed a pattern consistent with that result: participants rated feeling more successful ($\overline{x}=17.1$ versus $\overline{x}=15.1$). Taken together, these results indicate that these participants felt slightly more frustrated and strained by time while not quite as successful in achieving the task goals in the structured manual condition versus the interview; however, the overall did not feel particularly frustrated, time-constrained, or unsuccessful. Unequal-variances t-tests with Holm correction confirm these differences: temporal demand ($t = 3.09$, corrected $p = .0086$) and frustration ($t = 3.00$, corrected $p = .0086$) were significantly higher in the structured manual condition, while performance was significantly lower ($t = –4.80$, corrected $p < .0001$). No significant differences were observed for mental demand ($t = 0.75$, corrected $p = .907$) or effort ($t = 0.02$, corrected $p = .986$). Overall, these results indicate that although both conditions required similar mental effort, the structured manual increased perceived time pressure and frustration and reduced participants’ sense of task success relative to the elicitation interview.

The participant free responses corroborate these results: most indicated little to no struggle or strain in completing the task. As one participant described, ``the questions were straightforward, and did not take much mental demand when answering.'' However, some participants (though a minority) did have complaints about the slow response time from the LLM interviewer. Participants shared that they ``got aggravated at how long it took the interviewer to respond'' and found the process ``just a bit slow to go through.'' 

In contrast, while the structured manual condition took less time to complete, participants struggled more to figure out what to say. One described, ``It was difficult, I had to think for sometime before coming up with what to write.'' The uncertainty about what a successful specification looked like was a source of distress, with one participant remarking, ``I felt kind of annoyed and kind of frustrated about it. I also felt afraid that I was getting it wrong somehow. I didn't know what to write at first.'' However, similarly, the majority of participants had no complaints about the structured manual description experience.

\subsection{Structuring users' manual feed description rarely produced a strong feed difference}

The structured manual condition represents the capability of users to improve their own feed specifications if they are instructed at a high level on what to consider. In this condition, participants were given some guidance on what to discuss in their specifications, with reminders to think through content they want to exclude or prioritize, any intended use cases, and their desired experience, among other things. As noted in the reflections, though, users were not sure what they wanted or how to articulate it in this condition, even with the scaffolding.

While this intervention yielded modest but non-significant improvements in term of the post quality ratings, we overall do not see a pattern of improving feed quality. 
For many participants, the feeds were comparable in quality and experience. In fact, several noted that the feeds ``were similar in content and tone,'' or even ``felt like very much the same feed'' due to minimal changes made to the specification. Many participants noted the two feeds sharing multiple posts (or even a majority) or mentioned their inability to tell the feeds apart. A majority of participants did not have the motivation or ability to author a meaningfully different feed description even with instructions on information to include; they still struggled on how to articulate there preferences and therefore could not make strong improvements on their own.

\section{Discussion}

In this paper, we present \textit{feed elicitation interviews}, an LLM-powered interviewing system that helps users articulate what they want out of social media feeds and generate them from real posts. We find that through this elicitation process, users are able to produce feeds that they prefer to those that they manually describe, with minimal frustration or effort expended.

Here, we discuss the broader implications of this work, particularly the other opportunities for elicitation interviewing or similar systems, possible expansions of end-user social media feeds, the role of large language models in our system, and possible harms or abuses. We also discuss the limitations of both our system and the evaluation.

\subsection{The Design Space of Elicitation Tooling}

As our technological capacity expands and creation or tooling becomes available to non-technical end-users, more effort needs to be spent empowering people to figure out what they want to build and how. Practically, if we want to enable users to become their own designers, we should support them in the entire design process, rather than just building. Elicitation tools like ours can assist with these initial steps; interview-elicited feeds essentially help with problem definition and the early low-fidelity prototyping stages, helping users reflect on the potential results of their current design specification and make changes without having to fully implement the feed. Users can not only tune a feed, but figure out what kind of feed they want in the first place. 
Existing research on end-user tools shows that people benefit from these kinds of systems that help them gradually articulate and refine objectives that start out under-specified~\cite{ko2011state}. An interview allows a back and forth where intentions may be specified in increasing detail to resolve uncertainty. Grounding intentions like this is important, especially when interacting with tools with unconstrained input like natural language, in which case people choose the level of effort to dedicate towards developing and articulating their intentions~\cite{subramonyam2024gulf}. Our findings suggest that as LLMs become increasingly capable conversational partners, they can support preemptive personalization workflows, that allow them to appropriately specify their objective without excessive iteration with the end product.

Elicitation interviews open up possibilities well beyond social media feeds, because many systems require users to articulate intentions that are only partially formed at the outset. CHI research in personal informatics shows that people often enter with vague or unstable goals, and that technologies are more effective when they help users surface, clarify, and reflect on what they actually want to achieve~\cite{li2010stage,consolvo2009theory}. Similar dynamics appear in end-user automation, where users frequently specify simple trigger–action rules not because their goals are simple, but because these are the only aspects of their intentions they can easily express~\cite{ur2016trigger}. This mismatch between underlying intentions and representable form underscores the need for tools that help users articulate and refine their goals before system behavior is defined. Even in tools for behavior change, systems that support reflection and gradual goal formation are more successful than ones that assume users can fully articulate their intentions upfront~\cite{consolvo2009theory}. These results suggest a general pattern of use for elicitation processes like ours: when users need to decide on initial specifications for the system to take action on their behalf, and where iteration after the initial specification is undesirable due to its cost, whether in terms of programming or goal-pursuing effort.

Our approach to elicitation, however, still requires users to have an awareness of their needs or their desired use-cases for social media.  With ongoing advancements in contextually aware systems and user-modeling, there is a possibility to assist users in their self-needfinding with systems that help surface and suggest needs from observation of user behavior~\cite{shaikh2025gum}\, forming the initial basis for a feed or other artifact that the elicitation process helps to refine. In such hybrid systems, observed behaviors or needs could serve as a starting place for conversational refinement rather than something that the system interprets alone in the background. In fact, these models will likely be able to fill in some preference information themselves, especially as the model of a user's taste improves with time.

\subsection{Prescriptive Feed Design for User Well-being}

We see potential in directing custom feeds towards the well-being issues ascribed to social media. By having users prescriptively describe the content they want to see, this focuses them towards purposeful rather than passive use, which is associated with better mental and emotional well-being outcomes~\cite{verduyn2017social}. Feeds produced by reflective description could drive heightened mindfulness by focusing users on the reasons behind their social media use. While this space has yet to be fully explored, researchers have already built tools to allow users to preemptively express their desires for feed content, such as by prioritizing posts expressing their preferred values~\cite{kolluri2025alexandria} or controlling the rudeness or seriousness of posts~\cite{bhargava2019gobo}. An emerging vein of work aims to allow users the full range of expressivity in their feed creation, though the ideal user interaction design for feed specification to minimize user effort while maximizing agency is still undecided ~\cite{malki2025bonsai}.

At the same time, people can use their feed specifications to also avoid content that is upsetting, distracting, or otherwise distressing to them, which could otherwise drive negative mental health outcomes like anxiety, depression, or acute distress. Whereas social media platforms often have blanket restrictions on content that meet some bar of filtration for extremely toxic content, users have varying perspectives on what constitutes ``toxic'' or harmful for them, and may have different filtration needs. A custom feed constructor allows them to obstruct the content that they do not want, regardless of how close their preferences are to the rest of the population. Beyond interpersonal difference, people may have internal context-dependent perceptions of what makes content bothersome or distressing. Having control not just of \textit{what} they see, but also \textit{when} they see it, allows users to regulate their emotions and focus throughout their social media and lives. Existing work has already aimed to ameliorate user distress in this direction, for example, by allowing them to filter out content based on custom blocklists~\cite{jhaver2022filter} or downrank posts that express high levels of partisan animosity~\cite{jialam2024values, piccardi2024algorithms}.

We envision a potential future where people create a variety of purpose-driven feeds to address different purposes for social media use: informational, social, entertainment, etc. These feeds could be used individually, based on current use-case, or could be stitched together for a more varied browsing experience~\cite{liu2025decentralized}. Such an ecosystem allows for careful consideration of why and how to use social media, not just ahead of time, by specifying the feeds, but also in the moment, when choosing which feed(s) to peruse. Future work should investigate longitudinal impacts of social media use under such a paradigm, to see if it produces mental health or attentional improvements. 

\subsection{Using Large Language Models for Feed Elicitation and Creation Tools}

In its current state, our system relies heavily on a large language model (OpenAI's GPT-5) at all stages. Operations are therefore more expensive and time consuming than a more custom implementation: for example, the process of rating posts currently takes several minutes, which is prohibitively slow for a live, in-feed use case. While posts could be continuously rated asynchronously to be available whenever users arrived, this would increase the financial and computational cost. The time cost is primarily due to the initial, relevance filtering step. However, LLM classification of all posts is not the only way to implement retrieval. We see promise in a vector-based retrieval method, which could remove the need for LLM use in the initial step. There is also potential in fine-tuning smaller models on users' preferences to create a ranking model for the second step of post classification. 

Where the LLM's capacities are truly needed is throughout the interview, where the context-specific knowledge of different attributes and considerations was critical. For example, the LLM's ability to list potential subcategories of post topic or framing, as well as relevant emotions or use-cases, was incredibly useful, with participants frequently mirroring (or even copying and pasting) the suggestions and examples of the LLM interviewer, potentially lowering their manual (typing) and cognitive effort.

\subsection{Potential Harms and Abuses of End-User Feed Creation}

We envision a few potential directions in which harm could result from a wider facilitation of custom feed creation. We first consider the direction in which the system is used as intended towards antisocial or societally disadvantageous outcomes. For example, people could use feed controls to completely prevent their exposure to news, preventing them from being engaged or informed citizens.
People could also create extreme filter bubbles that prevent them from seeing any cross-cutting news and become completely embedded in a misinformation ecosystem. Perhaps, when given the option to completely control one's social media diet, people will only turn to the metaphorical ``candy'' and refuse to eat their ``vegetables.'' We envision a multi-feed paradigm, in which a user's social media ``diet'' draws from multiple feeds created to fulfill a variety of needs (social, informational, entertainment, etc.) that helps balance social media use across contexts or at least keep users mindful of their varying use types. A feed-creation system could itself suggest ``nutritious'' options (e.g., informational, lower in partisan animosity), or just help the user to explicitly consider those decisions rather than passively end up with an unbalanced diet. 

Outside of this ``in-line'' use, there are also possibilities to use feed creation for abuse or harassment. If feed creation is combined with publishing the feed to a broad audience, as the BlueSky ecosystem currently enables, feeds could be used to direct unwanted or negative attention towards vulnerable populations. In previous social media environments, accounts with small audiences had a reasonable expectation of anonymity, even if it was not guaranteed. Once feeds enable arbitrary feed creation, filtered from all public posts on the platform, this creates the opportunity for sudden visibility outside the scope of posters' expectations. This type of abuse has previously been facilitated by content reposting, like the LibsofTiktok account which has helped funnel hate and harassment, up to the level of bomb threats, to left-wing organizations~\cite{carless2023libs}. While combining targets into a public feed would not be the only vector for harassment campaigns, it is important to consider that such a path might allow for a broader set of victims including smaller accounts. We suggest a two-fold response to mitigate this risk: (1) educating users of BlueSky and other platforms with similar API access about the availability of public posts through the Firehose, and (2) an intermediate privacy setting for accounts that allows account browsing and following without additional permissions but prevents programmatic post access (as through the Firehose).

\subsection{Limitations}

\subsubsection{Where Elicitation Interviews Fail and Possible Mitigations}

While the elicitation interview does meaningfully improve post and feed quality for the user, several weaknesses still need to be mitigated.

One major problem is the system's deficiency at eliciting and representing prioritization. User prioritization of different factors can be extremely complex, with contingencies and hierarchical relationships determining what matters in different situations. Different types and strengths of user preferences are not always expressed clearly, much less correctly synthesized into the true importance hierarchy to appropriately rate posts according to user preference. An improved elicitation interview system could more directly elicit and represent the degree to which different preference details matter to the user, beyond our trinary rating system (though a complete representation of relative priorities would be extremely complex and prohibitively tiresome for a user to specify). The interview should also more explicitly communicate to the user that they can push back on questions if the specificity is getting irrelevantly high. 

Because the system currently relies on LLM calls for retrieval and ranking, the inventory is necessarily limited due to computational and time costs. However, if the retrieval were to be implemented differently, perhaps with vector similarity, a wider array of posts could be perused. A broader inventory could ameliorate the issue where the interview produces specifications that are too strict to find appropriate content.

Since we used a static set of posts and did not collect participant identity information, we were unable to implement certain feed ranking requests, like providing content specific to peoples' friends or family members. However, this is a common request, and relatively feasible to accomplish with network data tied to users' accounts. Future work should add that functionality. 

Our feed creation system was also not equipped to handle feed-level objectives like diversity or balance of different types of content. In cases where participants had a certain balance of content types in mind, we could not prioritize that goal in the feed creation. Feed-level objectives further complicate the relative prioritization conundrum discussed earlier, because a post-level classifier does not have the capability to prioritize some factors in some posts and not for others.
However, such objectives are plausible with a more complicated feed rating architecture and should be explored in future work.

\subsubsection{Limitations of the Evaluation}

Our study runs in a single session and we only measure immediate feed satisfaction. We therefore do not have a sense of how satisfaction may change over time. It could be that people's more specific preferences may be more variable, and therefore feed satisfaction would not stay stable after creation. A longitudinal study could measure the change over time, and also track mental health measures to investigate if prescriptively designed feeds benefit user well-being as we have speculated.

As mentioned above, each users' feeds are limited to using 10000-20000 posts found on BlueSky, which limits the available posts from what is technically available on the platform. A minimal number of participants were left with completely blank feeds due to this limit, and others received relatively inane content. However, we expect a larger inventory would particularly benefit the more specific feeds produced by the feed elicitation process.

Further, there are some limitations to the measurement accuracy. While we randomize the order of the feeds shown, users may have an idea of which is which, and base their ratings on factors beyond their actual preference (e.g., giving preferable ratings to the feed they worked harder on).

\subsection{Lessons for Preference Elicitation}

This project highlights practical lessons for designing preference elicitation systems in domains where users must specify complex or partially formed intentions, beyond social media feeds. Here, we focus on how those observations translate into concrete design implications.

First, elicitation systems should account for the fact that users’ preferences are often constructed during the interaction, not simply retrieved from memory or an underlying ground truth. In domains such as end-user automation, creative tools, or planning systems, people frequently refine or even change their goals once they see examples, suggestions, or trade-offs. Suggestions are therefore not merely aids for expressing pre-existing preferences, but mechanisms that actively shape preference formation. In our study, participants often adopted, edited, or reacted against options proposed by the system, using them as cognitive scaffolding to clarify what they did and did not want.

This suggests that elicitation interfaces should not rely on open-ended questioning alone, but should proactively introduce candidate dimensions, examples, and contrasts that users can respond to. Well-chosen suggestions can reduce cognitive effort, expose aspects of the design space users may not have considered, and make trade-offs more concrete. At the same time, because suggestions influence the preferences that emerge, systems should present them as provisional and revisable rather than authoritative defaults. Elicitation interfaces should include opportunities for users to revise earlier answers, compare alternatives, and react to concrete examples before the system commits to high-effort or high-stakes actions. Our system does allow revisions to the specification before feed creation for this reason, though a more sophisticated process could allow for further revision after viewing the artifact (feed) produced.

Second, elicitation tools should explicitly support partial or low-confidence preferences. As we observed, users will sometimes provide answers even when they do not feel strongly about a question, which can lead systems to overfit to weak signals. In other application areas—such as personalization in education or health contexts—this could produce inappropriate or overly rigid behavior. Interfaces can mitigate this by allowing users to mark preferences as tentative, low-priority, or “not sure,” and by treating such inputs differently in downstream decision-making. We implement a simple preference rating step, though further specificity and even explicit relative prioritization may be needed in other systems; such articulation can reach arbitrary levels of complexity depending on the needs of the elicitation process.

Third, systems should help users align their requests with what is actually feasible. When people describe ideal outcomes, they may unknowingly combine attributes that rarely co-occur or that the system cannot reliably produce. This issue is likely to arise in other settings involving recommendation, generative design, or automated planning. Elicitation tools can address this by showing representative examples, indicating when preferences are difficult to satisfy together, or suggesting nearby feasible alternatives rather than attempting to fulfill every request literally. Systems therefore ought to integrate an understanding of available capabilities or inventory into the elicitation process. This may be more straightforward when the inventory is static or slow to change (e.g., a library catalog) than when it is dynamic (e.g., social media content) or generated on demand (e.g., custom tool creation). LLMs already encode some contextual knowledge that can help estimate what outcomes are plausible, though this knowledge may not be sufficient to ensure that all preference combinations are feasible.

Together, these lessons suggest that preference elicitation in other domains could benefit from some of the structures we incorporate, like contextual suggestions and preference re-evaluation, with explicit handling of uncertainty and feasibility constraints, rather than a one-time request for a complete and internally consistent specification.

\section{Conclusion}

We introduce \textit{feed elicitation}, an LLM-powered interview-based method to help users carefully consider and articulate what they want out of a custom social media feed. In a design space dominated by tools to increase \textit{capacity} for expression, we aim to increase the \textit{fidelity} of users' expression.
The elicitation interview guides users through mindfully considering their intentions behind social media use and the posts that would serve them towards those intentions, as well as preemptively addressing undesired content.

We demonstrate the utility of this elicitation method by connecting it to a feed-generation system, creating an end-to-end feed creation system that takes users from first idea to a fully realized feed. We use this system to evaluate the efficacy of our elicitation process in specifying feeds that users prefer. We find that users prefer these interview-produced feeds to those that they manually describe and rate their content as more desired. This system therefore advances end-users' capabilities to control their own social media experiences towards ends they desire. Through feed elicitation, everyday social media users can create the custom social media feeds they desire---from learning about science to dreaming of vacation to keeping up with sports and media---even without design or engineering expertise.

\begin{acks}
We thank our anonymous reviewers as well as our colleagues Helena Vasconcelos, Yutong Zhang, and Michelle Lam for their thoughtful feedback on this manuscript. Our research was supported by NSF Award IIS-2403435, the Hoffman-Yee Research Grants at Stanford Institute for Human-Centered Artificial Intelligence (HAI), and the Stanford HAI-Hasso Plattner Institute Joint Research Program. Lindsay Popowski was supported by the Google PhD Research Fellowship and the Stanford Interdisciplinary Graduate Fellowship.  
\end{acks}

\bibliographystyle{ACM-Reference-Format}
\bibliography{bibliography}

@String{Computing = "Computing" }

@String{Computer = "{IEEE} Computer" }

@article{jialam2024values,
author = {Jia, Chenyan and Lam, Michelle S. and Mai, Minh Chau and Hancock, Jeffrey T. and Bernstein, Michael S.},
title = {Embedding Democratic Values into Social Media AIs via Societal Objective Functions},
year = {2024},
issue_date = {April 2024},
publisher = {Association for Computing Machinery},
address = {New York, NY, USA},
volume = {8},
number = {CSCW1},
url = {https://doi.org/10.1145/3641002},
doi = {10.1145/3641002},
journal = {Proc. ACM Hum.-Comput. Interact.},
month = apr,
articleno = {163},
numpages = {36}
}

@article{munn2020angry,
  title={Angry by design: toxic communication and technical architectures},
  author={Munn, Luke},
  journal={Humanities and Social Sciences Communications},
  volume={7},
  number={1},
  pages={1--11},
  year={2020},
  publisher={Palgrave}
}

@article{are2020instagram,
  title={How Instagram’s algorithm is censoring women and vulnerable users but helping online abusers},
  author={Are, Carolina},
  journal={Feminist media studies},
  volume={20},
  number={5},
  pages={741--744},
  year={2020},
  publisher={Taylor \& Francis}
}

@misc{kolluri2025alexandria,
title={Alexandria: A Library of Pluralistic Values for Realtime Re-Ranking of Social Media Feeds}, 
author={Akaash Kolluri and Renn Su and Farnaz Jahanbakhsh and Dora Zhao and Tiziano Piccardi and Michael S. Bernstein},
year={2025},
eprint={2505.10839},
archivePrefix={arXiv},
primaryClass={cs.HC},
url={https://arxiv.org/abs/2505.10839}, 
}

@article{jahanbakhsh2022peer,
author = {Jahanbakhsh, Farnaz and Zhang, Amy X. and Karger, David R.},
title = {Leveraging Structured Trusted-Peer Assessments to Combat Misinformation},
year = {2022},
issue_date = {November 2022},
publisher = {Association for Computing Machinery},
address = {New York, NY, USA},
volume = {6},
number = {CSCW2},
url = {https://doi.org/10.1145/3555637},
doi = {10.1145/3555637},
journal = {Proc. ACM Hum.-Comput. Interact.},
month = nov,
articleno = {524},
numpages = {40}
}

@inproceedings{mahar2018squadbox,
author = {Mahar, Kaitlin and Zhang, Amy X. and Karger, David},
title = {Squadbox: A Tool to Combat Email Harassment Using Friendsourced Moderation},
year = {2018},
isbn = {9781450356206},
publisher = {Association for Computing Machinery},
address = {New York, NY, USA},
url = {https://doi.org/10.1145/3173574.3174160},
doi = {10.1145/3173574.3174160},
booktitle = {Proceedings of the 2018 CHI Conference on Human Factors in Computing Systems},
pages = {1–13},
numpages = {13},
location = {Montreal QC, Canada},
series = {CHI '18}
}

@inproceedings{eslami2016like,
author = {Eslami, Motahhare and Karahalios, Karrie and Sandvig, Christian and Vaccaro, Kristen and Rickman, Aimee and Hamilton, Kevin and Kirlik, Alex},
title = {First I "like" it, then I hide it: Folk Theories of Social Feeds},
year = {2016},
isbn = {9781450333627},
publisher = {Association for Computing Machinery},
address = {New York, NY, USA},
url = {https://doi.org/10.1145/2858036.2858494},
doi = {10.1145/2858036.2858494},
booktitle = {Proceedings of the 2016 CHI Conference on Human Factors in Computing Systems},
pages = {2371–2382},
numpages = {12},
location = {San Jose, California, USA},
series = {CHI '16}
}

@inproceedings{eslami2015assumed,
author = {Eslami, Motahhare and Rickman, Aimee and Vaccaro, Kristen and Aleyasen, Amirhossein and Vuong, Andy and Karahalios, Karrie and Hamilton, Kevin and Sandvig, Christian},
title = {"I always assumed that I wasn't really that close to [her]": Reasoning about Invisible Algorithms in News Feeds},
year = {2015},
isbn = {9781450331456},
publisher = {Association for Computing Machinery},
address = {New York, NY, USA},
url = {https://doi.org/10.1145/2702123.2702556},
doi = {10.1145/2702123.2702556},
booktitle = {Proceedings of the 33rd Annual ACM Conference on Human Factors in Computing Systems},
pages = {153–162},
numpages = {10},
location = {Seoul, Republic of Korea},
series = {CHI '15}
}

@article{guess2023facebookstudy,
author = {Andrew M. Guess  and Neil Malhotra  and Jennifer Pan  and Pablo Barberá  and Hunt Allcott  and Taylor Brown  and Adriana Crespo-Tenorio  and Drew Dimmery  and Deen Freelon  and Matthew Gentzkow  and Sandra González-Bailón  and Edward Kennedy  and Young Mie Kim  and David Lazer  and Devra Moehler  and Brendan Nyhan  and Carlos Velasco Rivera  and Jaime Settle  and Daniel Robert Thomas  and Emily Thorson  and Rebekah Tromble  and Arjun Wilkins  and Magdalena Wojcieszak  and Beixian Xiong  and Chad Kiewiet de Jonge  and Annie Franco  and Winter Mason  and Natalie Jomini Stroud  and Joshua A. Tucker },
title = {How do social media feed algorithms affect attitudes and behavior in an election campaign?},
journal = {Science},
volume = {381},
number = {6656},
pages = {398-404},
year = {2023},
doi = {10.1126/science.abp9364},
URL = {https://www.science.org/doi/abs/10.1126/science.abp9364},
eprint = {https://www.science.org/doi/pdf/10.1126/science.abp9364},}

@inproceedings{bhargava2019gobo,
author = {Bhargava, Rahul and Chung, Anna and Gaikwad, Neil S. and Hope, Alexis and Jen, Dennis and Rubinovitz, Jasmin and Sald\'{\i}as-Fuentes, Bel\'{e}n and Zuckerman, Ethan},
title = {Gobo: A System for Exploring User Control of Invisible Algorithms in Social Media},
year = {2019},
isbn = {9781450366922},
publisher = {Association for Computing Machinery},
address = {New York, NY, USA},
url = {https://doi.org/10.1145/3311957.3359452},
doi = {10.1145/3311957.3359452},
booktitle = {Companion Publication of the 2019 Conference on Computer Supported Cooperative Work and Social Computing},
pages = {151–155},
numpages = {5},
location = {Austin, TX, USA},
series = {CSCW '19 Companion}
}

@inproceedings{wang2025authoring,
author = {Wang, Leijie and Yurechko, Kathryn and Dani, Pranati and Chen, Quan Ze and Zhang, Amy X.},
title = {End User Authoring of Personalized Content Classifiers: Comparing Example Labeling, Rule Writing, and LLM Prompting},
year = {2025},
isbn = {9798400713941},
publisher = {Association for Computing Machinery},
address = {New York, NY, USA},
url = {https://doi.org/10.1145/3706598.3713691},
doi = {10.1145/3706598.3713691},
booktitle = {Proceedings of the 2025 CHI Conference on Human Factors in Computing Systems},
articleno = {1214},
numpages = {21},
location = {
},
series = {CHI '25}
}

@article{han2024pressprotect,
author = {Han, Catherine and Li, Anne and Kumar, Deepak and Durumeric, Zakir},
title = {PressProtect: Helping Journalists Navigate Social Media in the Face of Online Harassment},
year = {2024},
issue_date = {November 2024},
publisher = {Association for Computing Machinery},
address = {New York, NY, USA},
volume = {8},
number = {CSCW2},
url = {https://doi.org/10.1145/3687048},
doi = {10.1145/3687048},
journal = {Proc. ACM Hum.-Comput. Interact.},
month = nov,
articleno = {509},
numpages = {34}
}

@inproceedings{jahanbakhsh2024browser,
author = {Jahanbakhsh, Farnaz and Karger, David R},
title = {A Browser Extension for in-place Signaling and Assessment of Misinformation},
year = {2024},
isbn = {9798400703300},
publisher = {Association for Computing Machinery},
address = {New York, NY, USA},
url = {https://doi.org/10.1145/3613904.3642473},
doi = {10.1145/3613904.3642473},
booktitle = {Proceedings of the 2024 CHI Conference on Human Factors in Computing Systems},
articleno = {946},
numpages = {21},
location = {Honolulu, HI, USA},
series = {CHI '24}
}

@article{stray2024building,
  title={Building human values into recommender systems: An interdisciplinary synthesis},
  author={Stray, Jonathan and Halevy, Alon and Assar, Parisa and Hadfield-Menell, Dylan and Boutilier, Craig and Ashar, Amar and Bakalar, Chloe and Beattie, Lex and Ekstrand, Michael and Leibowicz, Claire and others},
  journal={ACM Transactions on Recommender Systems},
  volume={2},
  number={3},
  pages={1--57},
  year={2024},
  publisher={ACM New York, NY}
}

@article{milli2025engagement,
  title={Engagement, user satisfaction, and the amplification of divisive content on social media},
  author={Milli, Smitha and Carroll, Micah and Wang, Yike and Pandey, Sashrika and Zhao, Sebastian and Dragan, Anca D},
  journal={PNAS nexus},
  volume={4},
  number={3},
  pages={pgaf062},
  year={2025},
  publisher={Oxford University Press US}
}

@inproceedings{subramonyam2024gulf,
author = {Subramonyam, Hari and Pea, Roy and Pondoc, Christopher and Agrawala, Maneesh and Seifert, Colleen},
title = {Bridging the Gulf of Envisioning: Cognitive Challenges in Prompt Based Interactions with LLMs},
year = {2024},
isbn = {9798400703300},
publisher = {Association for Computing Machinery},
address = {New York, NY, USA},
url = {https://doi.org/10.1145/3613904.3642754},
doi = {10.1145/3613904.3642754},
booktitle = {Proceedings of the 2024 CHI Conference on Human Factors in Computing Systems},
articleno = {1039},
numpages = {19},
location = {Honolulu, HI, USA},
series = {CHI '24}
}

@incollection{norman1986cognitive,
  title={Cognitive engineering},
  author={Norman, Donald A},
  booktitle={User centered system design},
  pages={31--62},
  year={1986},
  publisher={CRC Press}
}

@article{chandrasekharan2019crossmod,
author = {Chandrasekharan, Eshwar and Gandhi, Chaitrali and Mustelier, Matthew Wortley and Gilbert, Eric},
title = {Crossmod: A Cross-Community Learning-based System to Assist Reddit Moderators},
year = {2019},
issue_date = {November 2019},
publisher = {Association for Computing Machinery},
address = {New York, NY, USA},
volume = {3},
number = {CSCW},
url = {https://doi.org/10.1145/3359276},
doi = {10.1145/3359276},
journal = {Proc. ACM Hum.-Comput. Interact.},
month = nov,
articleno = {174},
numpages = {30}
}

@inproceedings{reddy2023rules,
author = {Reddy, Harita and Chandrasekharan, Eshwar},
title = {Evolution of Rules in Reddit Communities},
year = {2023},
isbn = {9798400701290},
publisher = {Association for Computing Machinery},
address = {New York, NY, USA},
url = {https://doi.org/10.1145/3584931.3606973},
doi = {10.1145/3584931.3606973},
booktitle = {Companion Publication of the 2023 Conference on Computer Supported Cooperative Work and Social Computing},
pages = {278–282},
numpages = {5},
location = {Minneapolis, MN, USA},
series = {CSCW '23 Companion}
}

@misc{piccardi2024algorithms,
      title={Social Media Algorithms Can Shape Affective Polarization via Exposure to Antidemocratic Attitudes and Partisan Animosity}, 
      author={Tiziano Piccardi and Martin Saveski and Chenyan Jia and Jeffrey T. Hancock and Jeanne L. Tsai and Michael Bernstein},
      year={2024},
      eprint={2411.14652},
      archivePrefix={arXiv},
      primaryClass={cs.CY},
      url={https://arxiv.org/abs/2411.14652}, 
}

@inproceedings{devito2017rip,
author = {DeVito, Michael Ann and Gergle, Darren and Birnholtz, Jeremy},
title = {"Algorithms ruin everything": \#RIPTwitter, Folk Theories, and Resistance to Algorithmic Change in Social Media},
year = {2017},
isbn = {9781450346559},
publisher = {Association for Computing Machinery},
address = {New York, NY, USA},
url = {https://doi.org/10.1145/3025453.3025659},
doi = {10.1145/3025453.3025659},
booktitle = {Proceedings of the 2017 CHI Conference on Human Factors in Computing Systems},
pages = {3163–3174},
numpages = {12},
location = {Denver, Colorado, USA},
series = {CHI '17}
}

@inproceedings{milton2023seeme,
author = {Milton, Ashlee and Ajmani, Leah and DeVito, Michael Ann and Chancellor, Stevie},
title = {“I See Me Here”: Mental Health Content, Community, and Algorithmic Curation on TikTok},
year = {2023},
isbn = {9781450394215},
publisher = {Association for Computing Machinery},
address = {New York, NY, USA},
url = {https://doi.org/10.1145/3544548.3581489},
doi = {10.1145/3544548.3581489},
booktitle = {Proceedings of the 2023 CHI Conference on Human Factors in Computing Systems},
articleno = {480},
numpages = {17},
location = {Hamburg, Germany},
series = {CHI '23}
}

@article{jansson1991fixation,
  title={Design fixation},
  author={Jansson, David G and Smith, Steven M},
  journal={Design studies},
  volume={12},
  number={1},
  pages={3--11},
  year={1991},
  publisher={Elsevier}
}

@article{moreno2016overcoming,
  title={Overcoming design fixation: Design by analogy studies and nonintuitive findings},
  author={Moreno, Diana P and Blessing, Luci{\"e}nne T and Yang, Maria C and Hern{\'a}ndez, Alberto A and Wood, Kristin L},
  journal={AI EDAM},
  volume={30},
  number={2},
  pages={185--199},
  year={2016},
  publisher={Cambridge University Press}
}

@incollection{moreno2015step,
  title={A step beyond to overcome design fixation: a design-by-analogy approach},
  author={Moreno, Diana P and Yang, Maria C and Hern{\'a}ndez, Alberto A and Linsey, Julie S and Wood, Kristin L},
  booktitle={Design Computing and Cognition'14},
  pages={607--624},
  year={2015},
  publisher={Springer}
}

@article{bakshy2015exposure,
  title={Exposure to ideologically diverse news and opinion on Facebook},
  author={Bakshy, Eytan and Messing, Solomon and Adamic, Lada A},
  journal={Science},
  volume={348},
  number={6239},
  pages={1130--1132},
  year={2015},
  publisher={American Association for the Advancement of Science}
}

@article{cinelli2021echo,
  title={The echo chamber effect on social media},
  author={Cinelli, Matteo and De Francisci Morales, Gianmarco and Galeazzi, Alessandro and Quattrociocchi, Walter and Starnini, Michele},
  journal={Proceedings of the national academy of sciences},
  volume={118},
  number={9},
  pages={e2023301118},
  year={2021},
  publisher={National Academy of Sciences}
}

@article{campana2017recommender,
  title={Recommender systems for online and mobile social networks: A survey},
  author={Campana, Mattia G and Delmastro, Franca},
  journal={Online Social Networks and Media},
  volume={3},
  pages={75--97},
  year={2017},
  publisher={Elsevier}
}

@article{rathje2021out,
  title={Out-group animosity drives engagement on social media},
  author={Rathje, Steve and Van Bavel, Jay J and Van Der Linden, Sander},
  journal={Proceedings of the national academy of sciences},
  volume={118},
  number={26},
  pages={e2024292118},
  year={2021},
  publisher={National Academy of Sciences}
}

@article{kubin2021role,
  title={The role of (social) media in political polarization: a systematic review},
  author={Kubin, Emily and Von Sikorski, Christian},
  journal={Annals of the International Communication Association},
  volume={45},
  number={3},
  pages={188--206},
  year={2021},
  publisher={Oxford University Press}
}

@article{allcott2020welfare,
  title={The welfare effects of social media},
  author={Allcott, Hunt and Braghieri, Luca and Eichmeyer, Sarah and Gentzkow, Matthew},
  journal={American economic review},
  volume={110},
  number={3},
  pages={629--676},
  year={2020},
  publisher={American Economic Association 2014 Broadway, Suite 305, Nashville, TN 37203}
}

@article{hunt2018no,
  title={No more FOMO: Limiting social media decreases loneliness and depression},
  author={Hunt, Melissa G and Marx, Rachel and Lipson, Courtney and Young, Jordyn},
  journal={Journal of Social and Clinical Psychology},
  volume={37},
  number={10},
  pages={751--768},
  year={2018},
  publisher={Guilford Press}
}

@article{shakya2017association,
  title={Association of Facebook use with compromised well-being: A longitudinal study},
  author={Shakya, Holly B and Christakis, Nicholas A},
  journal={American journal of epidemiology},
  volume={185},
  number={3},
  pages={203--211},
  year={2017},
  publisher={Oxford Academic}
}

@article{seabrook2016social,
  title={Social networking sites, depression, and anxiety: a systematic review},
  author={Seabrook, Elizabeth M and Kern, Margaret L and Rickard, Nikki S},
  journal={JMIR mental health},
  volume={3},
  number={4},
  pages={e5842},
  year={2016},
  publisher={JMIR Publications Inc., Toronto, Canada}
}

@article{ivie2020meta,
  title={A meta-analysis of the association between adolescent social media use and depressive symptoms},
  author={Ivie, Elizabeth J and Pettitt, Adam and Moses, Louis J and Allen, Nicholas B},
  journal={Journal of affective disorders},
  volume={275},
  pages={165--174},
  year={2020},
  publisher={Elsevier}
}

@article{shannon2022problematic,
  title={Problematic social media use in adolescents and young adults: systematic review and meta-analysis},
  author={Shannon, Holly and Bush, Katie and Villeneuve, Paul J and Hellemans, Kim GC and Guimond, Synthia},
  journal={JMIR mental health},
  volume={9},
  number={4},
  pages={e33450},
  year={2022},
  publisher={JMIR Publications Inc., Toronto, Canada}
}

@article{hormes2014craving,
  title={Craving F acebook? Behavioral addiction to online social networking and its association with emotion regulation deficits},
  author={Hormes, Julia M and Kearns, Brianna and Timko, C Alix},
  journal={Addiction},
  volume={109},
  number={12},
  pages={2079--2088},
  year={2014},
  publisher={Wiley Online Library}
}

@article{valenzuela2019paradox,
  title={The paradox of participation versus misinformation: Social media, political engagement, and the spread of misinformation},
  author={Valenzuela, Sebasti{\'a}n and Halpern, Daniel and Katz, James E and Miranda, Juan Pablo},
  journal={Digital Journalism},
  volume={7},
  number={6},
  pages={802--823},
  year={2019},
  publisher={Taylor \& Francis}
}

@article{allcott2017social,
  title={Social media and fake news in the 2016 election},
  author={Allcott, Hunt and Gentzkow, Matthew},
  journal={Journal of economic perspectives},
  volume={31},
  number={2},
  pages={211--236},
  year={2017},
  publisher={American Economic Association 2014 Broadway, Suite 305, Nashville, TN 37203-2418}
}

@article{lazer2018science,
  title={The science of fake news},
  author={Lazer, David MJ and Baum, Matthew A and Benkler, Yochai and Berinsky, Adam J and Greenhill, Kelly M and Menczer, Filippo and Metzger, Miriam J and Nyhan, Brendan and Pennycook, Gordon and Rothschild, David and others},
  journal={Science},
  volume={359},
  number={6380},
  pages={1094--1096},
  year={2018},
  publisher={American Association for the Advancement of Science}
}

@article{primack2017social,
  title={Social media use and perceived social isolation among young adults in the US},
  author={Primack, Brian A and Shensa, Ariel and Sidani, Jaime E and Whaite, Erin O and yi Lin, Liu and Rosen, Daniel and Colditz, Jason B and Radovic, Ana and Miller, Elizabeth},
  journal={American journal of preventive medicine},
  volume={53},
  number={1},
  pages={1--8},
  year={2017},
  publisher={Elsevier}
}

@article{bonsaksen2023associations,
  title={Associations between social media use and loneliness in a cross-national population: do motives for social media use matter?},
  author={Bonsaksen, Tore and Ruffolo, Mary and Price, Daicia and Leung, Janni and Thygesen, Hilde and Lamph, Gary and Kabelenga, Isaac and Geirdal, Amy {\O}stertun},
  journal={Health psychology and behavioral medicine},
  volume={11},
  number={1},
  pages={2158089},
  year={2023},
  publisher={Taylor \& Francis}
}

@misc{graze2025,
  author       = {Graze Social},
  title        = {Graze Social},
  year         = {2025},
  url          = {https://www.graze.social/},
  note         = {Accessed: 2025-09-01}
}

@misc{skyfeed2025,
  author       = {SkyFeed},
  title        = {SkyFeed},
  year         = {n.d.},
  url          = {https://skyfeed.app/},
  note         = {Accessed: 2025-09-01}
}

@misc{blueskyfeeds2025,
  author       = {BlueSky},
  title        = {Feeds - Bluesky},
  year         = {n.d.},
  url          = {https://bsky.app/feeds},
  note         = {Accessed: 2025-09-10}
}

@misc{blueskyfirehose2025,
  author       = {BlueSky},
  title        = {Firehose},
  year         = {n.d.},
  url          = {https://docs.bsky.app/docs/advanced-guides/firehose},
  note         = {Accessed: 2025-09-10}
}

@article{wartberg2021relevance,
  title={The relevance of emotion regulation, procrastination, and perceived stress for problematic social media use in a representative sample of children and adolescents},
  author={Wartberg, Lutz and Thomasius, Rainer and Paschke, Kerstin},
  journal={Computers in Human Behavior},
  volume={121},
  pages={106788},
  year={2021},
  publisher={Elsevier}
}

@article{carless2023libs,
  title={When Libs of TikTok posts, threats increasingly follow.},
  author={Carless, Will},
  journal={USA Today},
  pages={01A--01A},
  year={2023},
  publisher={USA Today}
}

@misc{shaikh2025gum,
  title={Creating General User Models from Computer Use}, 
  author={Omar Shaikh and Shardul Sapkota and Shan Rizvi and Eric Horvitz and Joon Sung Park and Diyi Yang and Michael S. Bernstein},
  year={2025},
  eprint={2505.10831},
  archivePrefix={arXiv},
  primaryClass={cs.HC},
  url={https://arxiv.org/abs/2505.10831}, 
}

@article{fardouly2015social,
  title={Social comparisons on social media: The impact of Facebook on young women's body image concerns and mood},
  author={Fardouly, Jasmine and Diedrichs, Phillippa C and Vartanian, Lenny R and Halliwell, Emma},
  journal={Body image},
  volume={13},
  pages={38--45},
  year={2015},
  publisher={Elsevier}
}

@article{holland2016systematic,
  title={A systematic review of the impact of the use of social networking sites on body image and disordered eating outcomes},
  author={Holland, Grace and Tiggemann, Marika},
  journal={Body image},
  volume={17},
  pages={100--110},
  year={2016},
  publisher={Elsevier}
}

@article{saiphoo2019meta,
  title={A meta-analytic review of the relationship between social media use and body image disturbance},
  author={Saiphoo, Alyssa N and Vahedi, Zahra},
  journal={Computers in human behavior},
  volume={101},
  pages={259--275},
  year={2019},
  publisher={Elsevier}
}

@article{rozenblit2002illusion,
  title={The misunderstood limits of folk science: An illusion of explanatory depth},
  author={Rozenblit, Leonid and Keil, Frank},
  journal={Cognitive science},
  volume={26},
  number={5},
  pages={521--562},
  year={2002},
  publisher={Wiley Online Library}
}

@article{kahneman1977intuitive,
  title={Intuitive prediction: Biases and corrective procedures},
  author={Kahneman, Daniel and Tversky, Amos},
  year={1977}
}

@book{suchman2007plans,
  title={Human-machine reconfigurations: Plans and situated actions},
  author={Suchman, Lucille Alice},
  year={2007},
  publisher={Cambridge university press}
}

@misc{schon1986reflective,
  title={The reflective practitioner: How professionals think in action},
  author={Schon, Donald A and DeSanctis, Vincent},
  year={1986},
  publisher={Taylor \& Francis}
}

@inproceedings{hsu2020awareness,
  title={Awareness, navigation, and use of feed control settings online},
  author={Hsu, Silas and Vaccaro, Kristen and Yue, Yin and Rickman, Aimee and Karahalios, Karrie},
  booktitle={Proceedings of the 2020 CHI Conference on Human Factors in Computing Systems},
  pages={1--13},
  year={2020}
}

@article{bernstein2023embedding,
  title={Embedding societal values into social media algorithms},
  author={Bernstein, Michael and Christin, Ang{\`e}le and Hancock, Jeffrey and Hashimoto, Tatsunori and Jia, Chenyan and Lam, Michelle and Meister, Nicole and Persily, Nathaniel and Piccardi, Tiziano and Saveski, Martin and others},
  journal={Journal of Online Trust and Safety},
  volume={2},
  number={1},
  year={2023}
}

@article{barnidge2017exposure,
  title={Exposure to political disagreement in social media versus face-to-face and anonymous online settings},
  author={Barnidge, Matthew},
  journal={Political communication},
  volume={34},
  number={2},
  pages={302--321},
  year={2017},
  publisher={Taylor \& Francis}
}

@article{levy2021social,
  title={Social media, news consumption, and polarization: Evidence from a field experiment},
  author={Levy, Ro’ee},
  journal={American economic review},
  volume={111},
  number={3},
  pages={831--870},
  year={2021},
  publisher={American Economic Association 2014 Broadway, Suite 305, Nashville, TN 37203}
}

@article{bail2018exposure,
  title={Exposure to opposing views on social media can increase political polarization},
  author={Bail, Christopher A and Argyle, Lisa P and Brown, Taylor W and Bumpus, John P and Chen, Haohan and Hunzaker, MB Fallin and Lee, Jaemin and Mann, Marcus and Merhout, Friedolin and Volfovsky, Alexander},
  journal={Proceedings of the National Academy of Sciences},
  volume={115},
  number={37},
  pages={9216--9221},
  year={2018},
  publisher={National Academy of Sciences}
}

@article{coyne2021tantrums,
  title={Tantrums, toddlers and technology: Temperament, media emotion regulation, and problematic media use in early childhood},
  author={Coyne, Sarah M and Shawcroft, Jane and Gale, Megan and Gentile, Douglas A and Etherington, Jordan T and Holmgren, Hailey and Stockdale, Laura},
  journal={Computers in Human Behavior},
  volume={120},
  pages={106762},
  year={2021},
  publisher={Elsevier}
}

@article{voinea2024digital,
  title={Digital slot machines: social media platforms as attentional scaffolds},
  author={Voinea, Cristina and Marin, Lavinia and Vic{\u{a}}, Constantin},
  journal={Topoi},
  volume={43},
  number={3},
  pages={685--695},
  year={2024},
  publisher={Springer}
}

@inproceedings{williams2025why,
author = {Williams, Gianna and Chen, Natalie and DeVito, Michael Ann and To, Alexandra},
title = {Why Can't Black Women Just Be?: Black Femme Content Creators Navigating Algorithmic Monoliths},
year = {2025},
isbn = {9798400713941},
publisher = {Association for Computing Machinery},
address = {New York, NY, USA},
url = {https://doi.org/10.1145/3706598.3713842},
doi = {10.1145/3706598.3713842},
booktitle = {Proceedings of the 2025 CHI Conference on Human Factors in Computing Systems},
articleno = {108},
numpages = {14},
location = {
},
series = {CHI '25}
}

@article{ma2022difference,
author = {Ma, Renkai and Kou, Yubo},
title = {"I'm not sure what difference is between their content and mine, other than the person itself": A Study of Fairness Perception of Content Moderation on YouTube},
year = {2022},
issue_date = {November 2022},
publisher = {Association for Computing Machinery},
address = {New York, NY, USA},
volume = {6},
number = {CSCW2},
url = {https://doi.org/10.1145/3555150},
doi = {10.1145/3555150},
journal = {Proc. ACM Hum.-Comput. Interact.},
month = nov,
articleno = {425},
numpages = {28}
}

@article{seering2024chillbot,
author = {Seering, Joseph and Khadka, Manas and Haghighi, Nava and Yang, Tanya and Xi, Zachary and Bernstein, Michael},
title = {Chillbot: Content Moderation in the Backchannel},
year = {2024},
issue_date = {November 2024},
publisher = {Association for Computing Machinery},
address = {New York, NY, USA},
volume = {8},
number = {CSCW2},
url = {https://doi.org/10.1145/3686941},
doi = {10.1145/3686941},
journal = {Proc. ACM Hum.-Comput. Interact.},
month = nov,
articleno = {402},
numpages = {26}
}

@article{leibmann2025redditrules, 
title={Reddit Rules and Rulers: Quantifying the Link Between Rules and Perceptions of Governance Across Thousands of Communities}, 
volume={19}, 
url={https://ojs.aaai.org/index.php/ICWSM/article/view/35863}, 
DOI={10.1609/icwsm.v19i1.35863}, 
number={1}, 
journal={Proceedings of the International AAAI Conference on Web and Social Media}, 
author={Leibmann, Leon and Weld, Galen and Zhang, Amy X. and Althoff, Tim}, 
year={2025}, 
month={Jun.}, 
pages={1098-1121} }

@article{verduyn2017social,
  title={Do social network sites enhance or undermine subjective well-being? A critical review},
  author={Verduyn, Philippe and Ybarra, Oscar and R{\'e}sibois, Maxime and Jonides, John and Kross, Ethan},
  journal={Social Issues and Policy Review},
  volume={11},
  number={1},
  pages={274--302},
  year={2017},
  publisher={Wiley Online Library}
}

@book{manning2008introduction,
  title={Introduction to information retrieval},
  author={Manning, Christopher D},
  year={2008},
  publisher={Syngress Publishing,}
}

@misc{malki2025bonsai,
      title={Bonsai: Intentional and Personalized Social Media Feeds}, 
      author={Omar El Malki and Marianne Aubin Le Quéré and Andrés Monroy-Hernández and Manoel Horta Ribeiro},
      year={2025},
      eprint={2509.10776},
      archivePrefix={arXiv},
      primaryClass={cs.HC},
      url={https://arxiv.org/abs/2509.10776}, 
}

@inproceedings{jhaver2022filter,
author = {Jhaver, Shagun and Chen, Quan Ze and Knauss, Detlef and Zhang, Amy X.},
title = {Designing Word Filter Tools for Creator-led Comment Moderation},
year = {2022},
isbn = {9781450391573},
publisher = {Association for Computing Machinery},
address = {New York, NY, USA},
url = {https://doi.org/10.1145/3491102.3517505},
doi = {10.1145/3491102.3517505},
booktitle = {Proceedings of the 2022 CHI Conference on Human Factors in Computing Systems},
articleno = {205},
numpages = {21},
location = {New Orleans, LA, USA},
series = {CHI '22}
}

@inproceedings{pommeranz2010user,
  title={User-centered design of preference elicitation interfaces for decision support},
  author={Pommeranz, Alina and Wiggers, Pascal and Jonker, Catholijn M},
  booktitle={Symposium of the Austrian HCI and Usability Engineering Group},
  pages={14--33},
  year={2010},
  organization={Springer}
}

@article{katz1969introduction,
  title={Introduction to reference work},
  author={Katz, William Armstrong},
  year={1969},
  publisher={McGraw-Hill}
}

@article{brown2008reference,
  title={The reference interview: theories and practice},
  author={Brown, Stephanie Willen},
  year={2008},
  journal={Published Works},
  url={https://digitalcommons.lib.uconn.edu/libr_pubs/13}
}

@article{mohedas2022designinterview, 
title={The use of recommended interviewing practices by novice engineering designers to elicit information during requirements development}, 
volume={8}, 
DOI={10.1017/dsj.2022.4}, 
journal={Design Science}, 
author={Mohedas, Ibrahim and Daly, Shanna R. and Loweth, Robert P. and Huynh, Linh and Cravens, Grace L. and Sienko, Kathleen H.},
year={2022}, 
pages={e16}
}

@article{ko2011state,
author = {Ko, Amy J. and Abraham, Robin and Beckwith, Laura and Blackwell, Alan and Burnett, Margaret and Erwig, Martin and Scaffidi, Chris and Lawrance, Joseph and Lieberman, Henry and Myers, Brad and Rosson, Mary Beth and Rothermel, Gregg and Shaw, Mary and Wiedenbeck, Susan},
title = {The state of the art in end-user software engineering},
year = {2011},
issue_date = {April 2011},
publisher = {Association for Computing Machinery},
address = {New York, NY, USA},
volume = {43},
number = {3},
issn = {0360-0300},
url = {https://doi.org/10.1145/1922649.1922658},
doi = {10.1145/1922649.1922658},
journal = {ACM Comput. Surv.},
month = apr,
articleno = {21},
numpages = {44}
}

@inproceedings{li2010stage,
author = {Li, Ian and Dey, Anind and Forlizzi, Jodi},
title = {A stage-based model of personal informatics systems},
year = {2010},
isbn = {9781605589299},
publisher = {Association for Computing Machinery},
address = {New York, NY, USA},
url = {https://doi.org/10.1145/1753326.1753409},
doi = {10.1145/1753326.1753409},
booktitle = {Proceedings of the SIGCHI Conference on Human Factors in Computing Systems},
pages = {557–566},
numpages = {10},
location = {Atlanta, Georgia, USA},
series = {CHI '10}
}

@inproceedings{consolvo2009theory,
author = {Consolvo, Sunny and McDonald, David W. and Landay, James A.},
title = {Theory-driven design strategies for technologies that support behavior change in everyday life},
year = {2009},
isbn = {9781605582467},
publisher = {Association for Computing Machinery},
address = {New York, NY, USA},
url = {https://doi.org/10.1145/1518701.1518766},
doi = {10.1145/1518701.1518766},
booktitle = {Proceedings of the SIGCHI Conference on Human Factors in Computing Systems},
pages = {405–414},
numpages = {10},
location = {Boston, MA, USA},
series = {CHI '09}
}

@inproceedings{ur2016trigger,
author = {Ur, Blase and Pak Yong Ho, Melwyn and Brawner, Stephen and Lee, Jiyun and Mennicken, Sarah and Picard, Noah and Schulze, Diane and Littman, Michael L.},
title = {Trigger-Action Programming in the Wild: An Analysis of 200,000 IFTTT Recipes},
year = {2016},
isbn = {9781450333627},
publisher = {Association for Computing Machinery},
address = {New York, NY, USA},
url = {https://doi.org/10.1145/2858036.2858556},
doi = {10.1145/2858036.2858556},
booktitle = {Proceedings of the 2016 CHI Conference on Human Factors in Computing Systems},
pages = {3227–3231},
numpages = {5},
location = {San Jose, California, USA},
series = {CHI '16}
}

@article{liu2025decentralized,
author = {Liu, Yuhan and Song, Emmy and Zhang, Owen Xingjian and Merriman, Jewel and Zhang, Lei and Monroy-Hern\'{a}ndez, Andr\'{e}s},
title = {Understanding Decentralized Social Feed Curation on Mastodon},
year = {2025},
issue_date = {November 2025},
publisher = {Association for Computing Machinery},
address = {New York, NY, USA},
volume = {9},
number = {7},
url = {https://doi.org/10.1145/3757688},
doi = {10.1145/3757688},
journal = {Proc. ACM Hum.-Comput. Interact.},
month = oct,
articleno = {CSCW507},
numpages = {25}
}

@article{Taylor1968QuestionNegotiationAI,
  title={Question-Negotiation and Information Seeking in Libraries},
  author={Robert S. Taylor},
  journal={Coll. Res. Libr.},
  year={1968},
  volume={76},
  pages={251-267},
  url={https://apps.dtic.mil/sti/html/tr/AD0659468/}
}

@article{belkin1982ask,
  title={ASK for information retrieval: Part I. Background and theory},
  author={Belkin, Nicholas J and Oddy, Robert N and Brooks, Helen M},
  journal={Journal of documentation},
  volume={38},
  number={2},
  pages={61--71},
  year={1982},
  publisher={MCB UP Ltd}
}


\appendix
\section{Appendix}

\subsection{Prompts} \label{app: prompts}

The following are the exact text of the prompts we use at various points of the interviewing, prompt-development, and post classification processes.

\subsubsection{Interviewer Prompt} \label{app: interviewer prompt}
\begin{verbatim}
You are an interviewer helping a person design a 
personalized social media feed of text posts sourced 
from Bluesky. You will conduct a brief interview, 
organized into short stages with different informational 
focuses, with the eventual goal of producing a 
specification (when prompted) for the feed that will 
be used to score individual posts for ranking. Ask 
questions relevant to the context of the user’s feed 
that are likely to uncover additional design insights 
and preferences without producing an excessively 
focused feed.
Craft each question towards understanding the current 
FOCUS while strictly following the INTERVIEWING 
PRINCIPLES.
\end{verbatim}

\subsubsection{Interviewing Principles} \label{app: interviewing principles}\hfill\\
\texttt{INTERVIEWING PRINCIPLES:
\begin{enumerate}
    \item Only ask one question at a time. Don’t combine multiple into one. 
    \item Ask questions that are easy to answer in one or two sentences.
    \item Stay focused on TEXT content for a SINGLE post (not visuals, balance of different types of posts, or metadata like author).
    \item Only suggest options that commonly appear in real Twitter/Bluesky posts and therefore could be found for the feed. Don't suggest niche topics or formats that are unlikely to appear on these platforms.
    \item When users are getting too specific to produce a feed, suggest ways to broaden while maintaining purpose.
    \item Make it clear that users do not have to give a preference if they don't have one. Avoid asking similar questions to those they didn't have a preference on.
    \item Aim for fewer than 4 questions per stage (ideal is 2–3).
\end{enumerate}}

\subsubsection{Interview Goals for Stage 1} \label{app: stage 1 goals}\hfill\\
\texttt{FOCUS: Figure out the purpose of the feed, particularly:
\begin{enumerate}
    \item Intention of the feed — What they want to get out of this feed, e.g., inspire, expose, inform, entertain, etc. How focused should it be to achieve this? 
    \item Relevant user context — What is the user’s current situation, interests, or needs that this feed should address (if feed content relies on this information)?
    \item Desired emotional or experiential outcome — How the user wants to feel or not feel
\end{enumerate}}

\subsubsection{Interview Goals for Stage 2} \label{app: stage 2 goals}\hfill\\
\texttt{FOCUS: Identify the relevant topics and content for the feed, especially:
\begin{enumerate}
    \item Additional topics/content — What peripheral topics should be included too?
    \item Topic focus — Which (if any) subtopics would they like to prioritize? Or is broadness and serendipity preferred?
    \item Undesired topics/content — Are there topics that appear on social media around this topic that would have a negative effect on the feed (distract\-ing or upsetting).
\end{enumerate}}

\begin{sloppypar}
\subsubsection{Interview Goals for Stage 3} \label{app: stage 3 goals}\hfill\\
\texttt{FOCUS: Identify what qualities posts should have to support the user goals.
\begin{enumerate}
    \item Post quality attributes — What qualities or characteristics should be prioritized in posts? Focus on qualities likely to be relevant in the given context (e.g., fact versus opinion for news, vocabulary level for science) and keep the options to those that are actually present on Twitter or Bluesky. 
    \item Exclusions — Are there qualities that would distress/distract/displease the user and should be avoided?
\end{enumerate}}
\end{sloppypar}

\subsubsection{Interview Goals for Stage 4} \label{app: stage 4 goals}\hfill\\
\texttt{FOCUS: Ensure there is no remaining important information the user wants to share but has not yet had a chance to specify.}

\subsubsection{Stage 5 Summary Generation Prompt} \label{app: stage 5 prompt}\hfill
\begin{verbatim}
Take this conversation and faithfully synthesize a 
concise description in the form below, where bracketed 
text [] is to be replaced. In synthesizing, respect and 
reflect user's different degree of preference (essential,
preferred, mildly preferred) and align your descriptions 
with real posts one can find on Bluesky:

[Overarching point and theme of the feed. What is going 
on at a high level, and what is the purpose? ]

Best types of post (0.8-1.0)
- [example class of post 1]
- [example class of post 2]... as needed

Desirable and decent: meets essential requirements but 
misses some other preferences (0.5-0.7)
- [1,2,...]

Acceptable: Doesn't meet all essential requirements but 
provides some value (0.3-0.5)
- [1,2...]

Better than nothing: Has some positive characteristics 
and nothing that is excluded (0.1-0.2)
- [1,2...]

Make sure to avoid (rate 0): [X]
Penalize posts for: [Y]

Rate posts that should be avoided or do not include any 
positive characteristics as 0.
\end{verbatim}

\begin{sloppypar}
\subsubsection{Reflection Prompt} \label{app: reflection prompt} \hfill\\
\texttt{
Evaluate whether the listed GOAL has been met in the conversation so far.
A GOAL is met only if the questions in the GOAL have been answered with clear, complete, and sufficiently 
specific information, unless the user explicitly denies to answer.
If the full answer to a question can be pieced together from multiple responses, consider that 
question answered.
If any answer is incomplete, vague, or requires further definition, consider that question not 
answered and the GOAL not met.
If the GOAL is met, output exactly: YES
If the user has indicated a desire to proceed or repeatedly has no preference, or the questions are 
becoming repetitive or overly specific, output exactly: YES
If the GOAL is not met, output exactly: NO
Do not provide explanations or reasoning.
}
    
\end{sloppypar}

\subsubsection{Classifier Creation Prompts} \label{app: classifier prompts}\hfill

The relevance description is created by prompting: 
\texttt{
    Write a *brief* paragraph describing what broad topics could be relevant to this feed. Topic means a high-level subject domain: what the post is about. Include both core and adjacent themes, not just those explicitly stated. Keep the topics high-level, WITHOUT qualifications or specifications. Do not include information about content type or framing.
}

\subsubsection{Classifier Execution Prompts}

The relevance score (binary) is classified as following, scoring each post in a batch of 10. This prompt structure was tested for fidelity to single prompt results and retained high agreement, significantly higher than a more simplistic version lacking the pseudocode structure.

\begin{verbatim}
Instructions:
    Run exactly the pseudocode below:
    ####
    initialize relevance_scores as empty list

    for i from 1 to 10:
        clear your memory of all posts and ignore relevance 
        scores
        hide all posts
        show only post[i]
        read eligibility criteria below

        if post[i] meets feed eligibility criteria:
            score = 1
        elif posts[i] partially meets feed eligibility 
        criteria but leaning towards relevance:
            score = 0.5
        elif posts[i] partially meets feed eligibility 
        criteria but leaning away from relevance:
            score = 0.4
        else:
            score = 0

        append score to relevance_scores

    output {{"relevance": relevance_scores}}
    Return only the JSON. Do not explain or comment 
    on any scores.
    ####

    # eligibility criteria: 
    # A post is eligible if it is relevant to something 
    in the following: [RELEVANCE CRITERIA]
    # stop reading here, refer to the posts below only 
    when pseudocode asks you to do so.
    # Posts:
\end{verbatim}

The quality classifier is created by appending the specification to the following classification instructions:

\texttt{Instructions: You are rating how appropriate a post is for the social media feed described above. Return your answer in the json object format: {{"quality": RATING}} where RATING is a number from 0.0 to 1.0 with one decimal point of precision. Post: }

\begin{sloppypar}
\subsubsection{NSFW Removal Prompt} \label{app: nsfw prompt} \hfill\\
\texttt{We are trying to identify posts that are obviously pornographic or linking to pornography. We are not trying to identify anything other than completely egregious content or that is explicitly indicated as nsfw.
Rank the following post either 0 (relatively acceptable) or 1 (completely inappropriate).
Give output of the json form {"nsfw": SCORE} with no explanations.
Post: }
\end{sloppypar}

{\subsection{Initial Study}\label{app: old study}

We conducted an earlier version of this study before we made system improvements. At a high level, from the version below to the system described in the main paper, we have changed the system prompts to aim for appropriate levels of specificity and ensure that suggested content types were likely to exist. The exact prompts for the past system are in Appendix~\ref{app: old prompts}. This older version did not include the step where users rate the importance of their different preferences to the interview. This older study was missing the questions about the elicitation experience, but included post ranking and employed a different rating scale and interface for the post rating. Below, we detail the study methodology, results, and analysis, as well as the old system prompts. 

\subsubsection{Participants}
We recruited $N=102$ participants from Prolific (\url{prolific.co}), an online crowdsourcing platform. We selected participants that had at least 50 submissions, were located in the United States, were native English speakers, had an approval rating of at least 95\%, had not completed our task before, and had some regular use of a text-and-feed-based social media (Facebook, Twitter, etc.). All participants received a payment of \$10 for approximately 40 minutes of their time, resulting in pay at a \$15 per hour rate.

\subsubsection{Procedure}

All participants started out on a page explaining the idea of feed creation. They are told that they will be constructing two versions of a feed designed to support them in some area of their social media use. They are specifically told that these feeds should not cover all topics they like to see, but a certain section, like entertainment, work, or news.

Their first action is to complete the baseline feed specification task. Then, they are randomly assigned to one of two enhanced elicitation methods, which they complete to create a second feed specification. Two different feeds---the baseline, and the treatment condition that they are assigned to---are generated according to the participant's preferences, which are randomly ordered and then presented, one at a time, to them.

When comparing the feeds, participants are first asked Likert-style questions about each of the twenty top posts in each feed (a total of forty posts). They are asked to rate their approval of that post on a 5-point scale (Disapprove - Neutral - Approve). After seeing both feeds, they are asked overall preference between the two on a seven-point Likert scale (Strongly Prefer Feed 1 - Prefer Feed 1 - Slightly Prefer Feed 1 - Neutral \ldots) and to explain their preference. They are then also asked to make twenty pairwise comparisons between posts in the two feeds to measure approval of the ranking and relative prioritization of different posts across the two feeds.

\subsubsection{Analysis Plan}

Through our previous analysis, we aimed to test whether feed elicitation selects content that users prefer to content in other feeds.
We also wanted to show that this content difference produced a difference in feed experience that registers to users. 

Our preference data came in three forms: overall feed preference (do they prefer the baseline or their assigned condition feed), the alignment of user pairwise post preference with each feed ranking, and the overall approval of each post in each feed. 

\paragraph{Feed-level Preference}

We analyzed feed preference ratings against the baseline condition using a Bayesian multinomial logit model with the categorical family in \texttt{brms}. This specification estimates the probability of each response category by condition without assuming parallel slopes across thresholds, making it appropriate for detecting non-uniform shifts and accommodating cases where some categories are sparse. Condition (structured manual versus elicitation interview) was included as a fixed effect, with random intercepts for participants to account for repeated measures:
\begin{small}
\begin{verbatim}
pref_rating ~ condition + (1 | participant_id)
\end{verbatim}
\end{small}
Because the categorical model treats each outcome category as distinct, it allows us to capture differences not only in overall tendencies toward one feed or the other but also in the likelihood of extreme versus moderate preferences across conditions.

Second, we conducted Wilcoxon signed-rank tests with continuity correction within each condition to test whether the median preference rating differed significantly from zero (representing neutral preference between feeds).

\paragraph{Post Quality}

We analyzed post-level approval ratings with a Bayesian cumulative logit mixed-effects model using the \texttt{brms} package, which allows category-specific effects of condition across thresholds within the Likert-style approval scale. This specification is necessary because our data violates the proportional odds assumption. The model included feed type (baseline, structured manual, elicitation interview) as a fixed effect, with random intercepts for participant and post: 
\begin{small}
\begin{verbatim}
approval ~ feed_type + (1 | participant_id) + (1 | post_id)
\end{verbatim}
\end{small}
After fitting the model, we conducted cutpoint-specific pairwise contrasts between the structured manual and elicitation interview conditions using \texttt{hypothesis()} in \texttt{brms}. This approach allows us to assess both absolute improvements over the baseline and the relative advantage of the interview feed compared to the structured manual feed in terms of the user-assessed quality of different posts.

We analyzed pairwise choice alignment by fitting a logistic regression model predicting whether the higher-ranked post in a pair was selected by the participant, in the pairwise comparisons. We used both the ranking of those posts in the manual feed and the interaction between assigned condition and ranking in that condition's feed as predictors. This let us test whether alignment with the treatment feed differs by condition (interview versus structured manual):
\begin{small}
\begin{verbatim}
upranked ~ baseline_agrees + condition_agrees 
           + condition_agrees:condition + (1 | participant_id)
\end{verbatim}
\end{small}
Here, baseline\_agrees reflects whether the baseline ranks the chosen post higher, condition\_agrees the condition feed's agreement, and condition whether the participant was in the elicitation interview or structured manual group. We can test whether the treatment feeds outperform the manual, and whether that outperformance is larger in one condition than the other. This measure assesses how well the feeds represent the participant's own prioritization of different posts across the two feeds.

\paragraph{Qualitative Analysis of Specifications and Interviews}

In addition to these quantitative measures, we looked to the users' written feedback about the feeds they saw to contextualize their preference. We performed inductive coding on the different explanations of user preference to understand the major reasons users preferred one feed over another. We followed by lightly coding the elicitation interviews to understand possible failure or success cases of the interview. We looked for signs of user frustration or satisfaction, and then tried to understand the patterns of behavior that produced these reactions. We also looked for questions that prompted responses that produced key outcomes in the eventual specification.

We also analyzed themes in the intermediate artifacts produced by our participants: the feed specifications, which we hand code. We looked for characteristics either common in feed specifications or that we aimed to produce via the specification process: general desire for feed experience, topic(s) of focus, topics to avoid, content characteristics to avoid, characteristics of desired content. Here, we aimed to see if the interview had the desired impact in prompting users to consider and articulate their feed preferences comprehensively.  

\subsubsection{Results}
\paragraph{Participants preferred content resulting from elicitation interviews}

Participants rated posts higher at all levels in the elicitation condition than the baseline, and were more likely to rate posts at the Approve level (vs. Neutral) in the elicitation condition compared to the structured manual condition.

Though all conditions had more posts approved than disapproved, the elicitation interview had a higher proportion of approved posts (67\% versus 60\% for baseline and structured manual) and fewer disapproved posts (13\% versus 20\% for the baseline and 18\% for structured manual). This corresponds to a reduction by over a third of disliked content and an increase of over 10\% of approved content versus the baseline. Exact breakdowns of the approval rates are shown in Figure~\ref{fig: approval distribution}.

\begin{figure}[tb]
  \centering
    \includegraphics[width=\columnwidth]{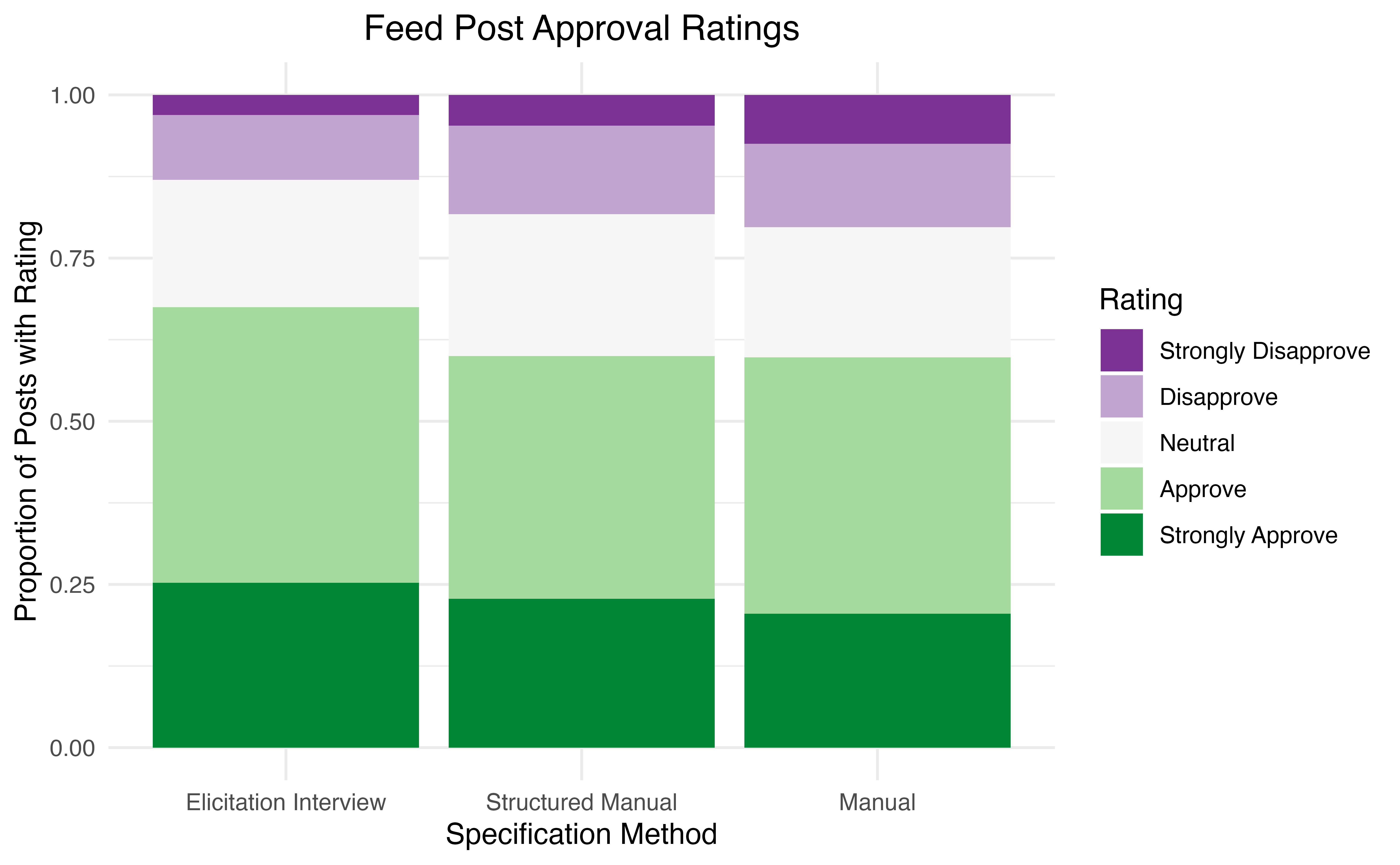}
  \caption{Participants approved the posts at a modest but significantly higher rate for feeds created by the purpose interview, with 67\% rated as approve or strongly approve versus 60\% for the baseline and structured manual control. }
  \label{fig: approval distribution}
\end{figure}

We tested this result by fitting a cumulative logit mixed model with category-specific effects of condition with \texttt{brms}, allowing treatment effects to vary across thresholds in the rating scale (with random effects for post and participant). Each estimate describes the relative likelihood of a rating one higher (e.g. Approve rather than Neutral).

Relative to the baseline, the elicitation interview increased the log-odds of higher ratings across all four thresholds, with odds ratios ranging from $\sim$1.5-2. In each case, the $95\%$ credible interval excluded zero, indicating strong posterior evidence that the elicitation interview reliably increased the probability of assigning higher ratings throughout the entire scale. The effect was strongest at the bottom of the scale, meaning that the elicitation interview reduced the likelihood of the very lowest rating the most, though the likelihood of higher ratings also increases by a factor of around $1.5$.
By contrast, the structured manual condition showed more limited effects. The first two thresholds showed positive effects with 95\% credible intervals excluding zero while the latter two intervals overlapped zero, suggesting little to no posterior evidence for an effect at the upper thresholds. All the estimates for each threshold are reported in Table~\ref{tab: post approval model}. 

To directly compare the structured manual and elicitation interview conditions, we tested threshold-specific contrasts of their coefficients using Bayesian linear hypothesis tests. At the first two thresholds (Strongly Disapprove to Disapprove, Disapprove to Neutral), the estimated difference was small and provided no evidence of divergence between conditions. At the Neutral to Approve threshold, however, the elicitation interview condition showed a clear advantage over structured manual, with a notable positive difference of $0.36$ (95\% credible interval excluded zero). At the final threshold (Approve to Strongly Approve), the difference favored the elicitation interview condition but was not notable. Exact intervals are reported in Table~\ref{tab: post approval model}.

All together, the model results suggested that both the structured manual and feed elicitation conditions reliably decreased the odds of posts receiving the lowest quality ratings, relative to the baseline feed. However, only the feed elicitation condition maintained positive effects at higher thresholds, with posterior estimates consistently favoring higher ratings even into the top categories.

\begin{table*}[t]
\centering
\renewcommand{\arraystretch}{1.4}
\begin{tabular}{lcccc}
\hline
 & Strongly Disapprove $\rightarrow$  
 & Disapprove $\rightarrow$  
 & Neutral $\rightarrow$  
 & Approve $\rightarrow$  \\
 & Disapprove
 & Neutral
 & Approve
 & Strongly Approve \\
\hline
\multicolumn{5}{l}{\textit{Elicitation Interview vs. Baseline}} \\
Log-odds estimate 
& \textbf{0.72} [0.24,\,1.22]  & 
\textbf{0.51} [0.22,\,0.80]  & 
\textbf{0.37} [0.14,\,0.61] & 
\textbf{0.42} [0.16,\,0.68] \\
Odds Ratio 
& 2.06 [1.27,\,3.39] &
1.66 [1.25,\,2.22] &
1.45 [1.15,\,1.84] &
1.52 [1.17,\,1.98] \\
\\
\multicolumn{5}{l}{\textit{Structured Manual vs. Baseline}} \\
Log-odds estimate &
\textbf{0.84} [0.43,\,1.27]  & 
\textbf{0.26} [0.01,\,0.52] & 
0.01 [$-$0.21,\,0.22] & 
0.13 [$-$0.12,\,0.37] \\
Odds Ratio &
2.32 [1.54,\,3.56] &
1.30 [1.01,\,1.68] &
1.01 [0.81,\,1.25] &
1.14 [0.89,\,1.44] \\
\\
\multicolumn{5}{l}{\textit{Pairwise Contrasts: Elicitation Interview vs. Structured Manual}} \\
Log-odds difference &
$-$0.12 [$-$0.71,\,0.49] &
0.24 [$-$0.12,\,0.61] &
\textbf{0.36} [0.07,\,0.67] &
0.29 [$-$0.04,\,0.62] \\
\\
\hline
\\[-1.8ex]
\end{tabular}
\caption{\textbf{Category-specific effects of condition versus the standard baseline} from the cumulative logit mixed model, with an additional block reporting \textit{pairwise contrasts} between the two non-baseline conditions, computed with \texttt{brms::hypothesis}. We bold notable (95\% credible intervals excluding zero) effects. The elicitation interview shows consistently higher odds of giving better ratings across thresholds, while structured manual mainly reduces the odds of the lowest ratings. Direct contrasts indicate that the interview condition provides stronger upward shifts than structured manual at the upper thresholds.}
\label{tab: post approval model}
\end{table*}

In terms of ranking, the feeds produced by the elicitation interview were the most consistent with the pairwise rankings performed by the participants, agreeing with 61\% of the pairwise ranks, versus 51\% agreement for the structured manual specification and 48\% for the baseline.\footnote{Agreement means that the post ranked higher by the participant is ranked higher by that feed. If both posts are not found in the feed, the absent post is considered to be ranked lower.} Our statistical model supports this finding.

\begin{table*}[!ht] 
\centering
\renewcommand{\arraystretch}{1.4}
\begin{tabular}{@{\extracolsep{5pt}}lc} 
\hline 
& \textit{Dependent variable: Top post ranked} \\ 
\hline 
Constant & $-$0.290$^{**}$ (0.109) \\ 
Baseline feed agrees & 0.170 (0.111) \\ 
Condition feed agrees & 0.366$^{**}$ (0.119) \\ 
Condition feed agrees : Elicitation Interview Condition & 0.445$^{**}$ (0.150)\\ 
\hline
\\[-1.8ex]
\end{tabular} 

\caption{\textbf{Results of a mixed-effects logistic regression to predict pairwise rankings.} Participant rankings are significantly predicted by the rankings of the structured manual and elicitation interview conditions, but the elicitation interview is an even stronger predictor. Note that $^{*}$p$<$0.05; $^{**}$p$<$0.01; $^{***}$p$<$0.001.} 
\label{} 
\end{table*} 

The logistic mixed-effects regression shows that the coefficient for the feed’s rank difference is positive but low ($0.170$, corresponding to an odds ratio of 1.2) and non-significant ($p = 0.13$). This indicates that the baseline feed ranking does not align meaningfully with the participants' own prioritization of posts. There are positive and significant coefficients for the agreement of either condition feed: $0.366$ ($p= 0.0021$), \textit{and} for the interaction between condition feed agreement and the condition being the interview: $0.445$ ($p=0.0030$). We interpreted this to mean that either condition feed agreeing with a ranking makes that ranking 1.44 times more likely, but if the condition was the elicitation interview, the likelihood increases again by a factor of 1.56, for a odds ratio of 2.25 in total. More importantly, not only did the elicitation interview outperform the baseline in terms of aligning with user preference, it also outperformed the structured manual specification. 

Between these two analyses, we could tell that participants preferred content produced by the elicitation interview to that from both the baseline and structured manual specification methods. This partiality holds in terms of independent per-post ratings and between-post preference data. Qualitative feedback reflected improvements in post preference as well. Participants praised the elicitation interview feed for being ``closer to what I described,'' ``exactly how I wanted it to be for the most part,'' and ``much more relevant to my interests.'' We turned to analysis of between-feed preference data to see if the post-wise quality improvement consistently translated into a difference in perceived quality of the overall feed.

\paragraph{Interview-elicited feeds provided high-risk but high reward}

When asked overall feed preference, participants preferred the interview-elicited feed over the baseline 40\% of the time, and preferred the baseline over it 38\% of the time---this corresponds to a 51-49 split when users communicated a preference one way or another. Only 22\% of the time were participants neutral between the baseline and interview-elicited feed.
In contrast, participants preferred the structured manual feed 31\% of the time, preferred the baseline over it 25\% of the time, and had no preference 44\% of the time, twice the rate of neutrality as compared to interview elicitation. The strongest difference between the elicitation interview and the structured manual specification, it seems, is a tendency for more polarized reactions, rather than a strict tendency towards preference in one direction across all participants. The median preference level, in contrast, was neutral (zero preference) for either condition. For both the elicitation interview and the structured manual condition, Wilcoxon signed-rank tests (with continuity correction) did not reveal a significant deviation from neutral preference with respect to the baseline specification (elicitation interview: median 0, $V = 460$, $p = 0.32$; structured manual: median 0, $V = 266$, $p = 0.29$). 

\begin{figure}[tb]
  \centering
    \includegraphics[width=\columnwidth]{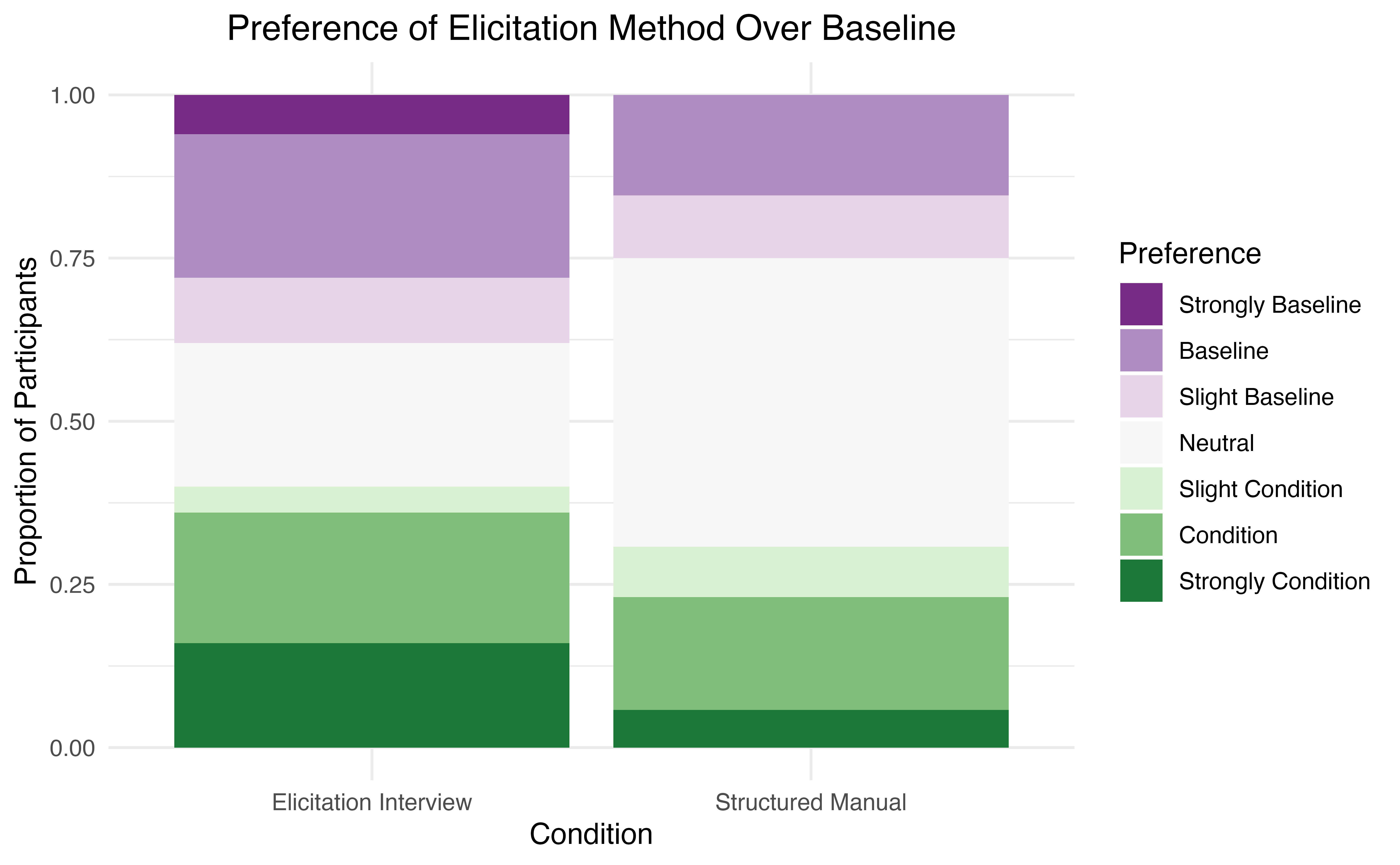}
  \caption{Participants overall preferred both the elicitation interview and the structured manual condition feeds over the baseline. However, participants who completed the interview leaned towards more extreme preferences versus the baseline, feeling strong preferences for and against the interview condition at higher rates. Over a third of participants had no preference between the two manual specifications, whereas less than a fifth had no preference between the interview and baseline feeds.}
  \label{fig: preference distribution}
\end{figure}
\begin{table*}[ht]
\centering
\renewcommand{\arraystretch}{1.4}
\begin{tabular}{lcc}
\hline
Preference Expression (vs.\ Neutral) & Log-odds (Interview vs.\ Structured Manual) & 95\% CrI \\
\hline
Strongly prefer baseline ($-3$)  & \textbf{7.14} & [0.98, \, 22.54] \\
Prefer baseline ($-2$)           & \textbf{1.14} & [0.01, \, 2.32] \\
Slightly prefer baseline ($-1$)  & 0.80 & [$-0.72$, \, 2.27] \\
Slightly prefer condition ($+1$) & 0.04 & [$-1.96$, \, 1.92] \\
Prefer condition ($+2$)          & 0.92 & [$-0.20$, \, 2.09] \\
Strongly prefer condition ($+3$) & \textbf{1.85} & [0.38, \, 3.52] \\
\hline
\\[-1.8ex]
\end{tabular}
\caption{Bayesian multinomial logit model (categorical family) comparing preference (versus baseline) of the elicitation interview to that of structured manual. Coefficients represent log-odds of selecting each response category relative to Neutral. We bold notable (95\% credible intervals excluding zero) effects. There is strong evidence for increased likelihood of extreme preferences---for and against---with the elicitation interview versus the structured manual method.}
\label{tab: overall pref}
\end{table*}

We used a Bayesian multinomial logit model (categorical family in \texttt{brms}) to estimate the probability of each rating category by condition. This approach does not assume proportional odds and is equipped to model the observed polarization pattern. 
Compared to the structured manual condition, the elicitation interview condition receives notably more extreme responses. The odds of choosing ``Strongly prefer condition'' vs ``No preference'' were increased (log-odds $1.9$) as were the odds of ``Strongly prefer baseline'' (log-odds 7.4).\footnote{The extreme value here is likely because no participants indicated that they strongly preferred the baseline feed over the structured manual feed.} Middle categories (slight preference either way) did not differ reliably. Full model results are reported in Table~\ref{tab: overall pref}. This model therefore gives us evidence that the elicitation interview produced stronger preference between feeds versus the structured manual specification. Participants did not overall prefer or reject the interview feeds more than the structured manual feeds, but they held extreme preferences at higher rates: when they liked it, they were likely to strongly like it, and when they disliked it, they were likely hold a strong dislike.

\paragraph{Participants appreciated additional specificity and personalization in feeds produced via interview elicitation}

When participants had distinct preferences that they failed to communicate, they really appreciated the difference in their elicited feeds. One participant described the difference:
\begin{quote}
    Feed One felt closer to what I described — thoughtful, informative, and curiosity-driven posts (space science, creativity, authentic reflections, puzzles/learning). It leaned more toward the blend of learning + inspiration + enjoyment I wanted. Feed Two had some interesting items, but also more filler and things that felt tangential.
\end{quote}

Other participants remarked on how much the interview personalized to them, noticing that the feeds were ``more aligned with my core interests in science, equity, and global issues,'' or ``more authentic and tailored to my interests.'' In these cases, the baseline feed sometimes served them content that felt like obvious mistakes, like having ``weather that wasn't relevant and biased Ukraine/Russia news I don't want to see.'' Participants frequently left critical requirements unstated in their initial specifications, like stating that they wanted ``a lot of sports posts that are positive and make me excited'' and then preferring a feed that served ``informational posts that provided scores and more clear information'' versus ``others oponions and thoughts more [sic].'' The interview also helped participants remember to omit relevant but disruptive content, like ``posts that are profanity-heavy or feel performatively edgy,'' ``climate doomism, culture-war hot takes, and clickbait content,'' ``celebrity gossips,'' ``ADHD/health advice threads,'' ``spoilers from Movies and Tv shows,'' and ``influencer marketing junk.'' These omissions produced notable quality improvements, like removing offensive content on a vacation experience-oriented feed. The full process of making these specification adjustments is shown in Table~\ref{tab:interview-improvements-old}, which starts with the initial feed description, and shows how a particular interview question resulted in a positive change for the feed.

These observations also align with the quantitative rating data, where the elicitation interview produced more ``good'' content and less ``bad,'' both tailoring to their interests and removing the content that they had not thought to rule out.

 \newcolumntype{L}[1]{>{\raggedright\arraybackslash}p{#1}}
 
\begin{table*}
  \centering
  \renewcommand{\arraystretch}{1.6}
  \small
  \begin{tabular}{L{1.3in} L{1.1in} L{1.2in} L{1.1in} L{0.9in}}
  \hline
    Initial Specification & Question & Response & Specification & User Reflection \\
    \hline
    \ldots The content I like is usually something that someone made or a story or a good quality article. There is a lot of repetitive content made for views that I don't like from everyone, only specific users. I also don't like to see sentimental or too much celebrity content.
    & When you’re winding down, which topics do you most enjoy in posts or articles? For example: nature/animals, space, everyday science, design/architecture, history tidbits, tech explained simply, travel/culture \ldots
    & I like posts about film and technical things like DIY electronics or explanations of larger scale data networks and things like that. I also like learning random things particularly about space or a deeper dive into some historical or geological thing.
    & Prioritize original, maker-style posts and thoughtful stories where the creator interprets and explains how things work \ldots Include real science updates and digestible deep dives on space, history, and geology \ldots
    & [The feed] had more content I could learn from
    \\
    I want to see a lot of sports posts that are positive and make me excited. I also want to see movie and TV posts that share interesting news about upcoming releases and things coming out soon.
    & For sports posts, which positive themes do you want most—e.g., big clutch highlights, inspiring comeback stories, wholesome team moments, acts of sportsmanship, funny/heartwarming behind-the-scenes, or milestone achievements?
    & I like Highlights the most and score updates as well. Achievements could be scattered throughout as well
    & Rank highest: ultra-brief sports highlights and final scores from the NBA, NFL, UFC, and tennis
    & [It had] informational posts that provided scores and more clear information while the second featured others oponions and thoughts more. [sic] \\
    
    I want to see accurate information \ldots I want to see uplifting news as well as real news to balance out my feed when it comes to real life news or politics. I want to see animal videos \ldots I love to see how our technology, psychology, and facts about Earth and our own bodies are changing as we move into the future \ldots some culture related facts in other countries or live feeds of cultural events currently happening would be a fun addition.
    & What tone should posts use to best fit your goals? (Examples to consider: calm/neutral explainers, hopeful/uplifting framing, curiosity-driven “what we know/what we don’t,” lightly humorous without snark, or empathetic when covering tough news.)
    & Empathetic when it's tough, yes. I would like comedy as long as it doesn't cross the lines into mean spirited or bigoted. I don't want any of what they call on the internet \"doomer\" mentality. I want hopeful and uplifting posts, motivating us to take action rather than fearing us back into our homes if you get what i mean.
    & Tone should be empathetic on hard news and hopeful or lightly humorous without mockery; avoid “doomer” fatalism.
    & [The feed] felt more hopeful, there was more balance between serious topics and optimistic topics\\
    &&&& \\

    \hline
    \\[-1.8ex]
    \end{tabular}
    \caption{The interview gives users a chance to clarify underspecified points from their original specification. These clarifications had significant impacts in the resulting feeds and produced qualitatively better browsing experiences for the participants.}
    \label{tab:interview-improvements-old}
\end{table*}

\paragraph{Increased specificity from elicitation can also worsen feeds}

Specificity was not a purely beneficial characteristic, however. We found evidence to suggest that overly expressive specifications could harm the overall feed experience in several ways. A noisy association holds between the level of initial specificity in topic and preference towards the baseline feed over the interview one. For example, ``I'd like to see posts about new science articles like physics, space/cosmology, \& AI breakthroughs/discoveries. I'd also love seeing stuff from comedians like Big Jay Oakerson, Mark Normand, Dan Soder, Joe List, Joe DeRosa, and other comedians,'' versus ``I want to see posts about music, art, food, and nature.'' When we analyzed the interview transcripts for participants with a strong negative preference (excluding those who only slightly preferred the baseline) for the interview versus baseline feed, we noticed a pattern of overspecification, namely including relatively unimportant or irrelevant preferences. Many of the participants who ended up preferring the interview feed, inversely, adeptly communicated their lack of preference on questions that did not interest them. 

Multiple participants critiqued the interview-produced feed for not emphasizing a particular factor that mattered to them that was emphasized in their original specification. For instance, one specified that ``I'd like to see more positive posts to enjoy my life. I love reading posts helping me to grow and learn more. They can be about discovering new places and interesting ideas.'' Their interview specification instead focused on learning about fashion, but they ended up preferring the baseline feed because ``it had more positive posts included.'' By adding more factors and preference details, the interview effectively diluted what they most cared about. This same pattern emerged for a participant who most cared about seeing personal announcements like weddings, but saw that content underemphasized in their interview feed. 

As implemented, our elicitation interview tended to focus toward ever-more-specific feed ideas, without a principled stopping point. Its tendency towards overspecification does not just dilute the signal of what users truly care about, but also may rule so many posts as to leave only low-quality content. The most common complaint against a elicitation interview feed was that some of the posts were irrelevant and not what they wanted. As one participant explained, ``[The baseline feed] had clearer and more useful posts. [The elicitation interview feed] had some I didn’t care about.'' While the specification holds some of the blame, the available inventory of posts shares responsibility. In half of cases where participants flagged this issue (4/8), the posts chosen for the elicitation interview feed are markedly lower-rated than the posts for the baseline feed, indicating that there were insufficient posts up to the standard of the interview specification. A quarter more had small but apparent differences in ranking score between the feeds. This complaint highlights a major downfall of our system: an inability to set expectations for participants about what feeds are feasible based on the inventory. 

We attempted to mitigate this in the interview: direct instructions to accept broad user preferences, limits on the number of questions asked, and a relevance rating step that deliberately broadens the scope to identify potential posts. However, the interview was not grounded in the available inventory of posts and could therefore promise a user the ability to select incredibly niche posts that match their wants exactly, only for that specification to rule out many posts they would have been satisfied with.
An improved system could either search through a larger set of posts or give feedback to the user when it failed to retrieve significant quality content.

The elicitation interview can benefit people who have left out critical details of what they want from their feed. However, the appropriate specificity of a feed descriptions is hard to measure, because it depends significantly on what the user truly prioritizes

\paragraph{Elicitation interviews struggle to establish relative priority of preferences}

If users' preferences were truly, completely translated into the specification and ranking system, being overly specific would not be an issue, because the content closest to what the user wanted would be served to them. We believe that the key issue in the feeds that under-delivered for users was that the different preferences were not appropriately weighed against each other.

Participants overwhelmingly added new preferences and requirements to their specifications and approved them, indicating that the preferences elicited were not strictly false. We take this as evidence that some of their preferences were initially unexpressed in a manual specification scenario. However, since the resulting feeds were not preferred, there must be a gap in successfully translating user preference data into actual feed rating and ranking. As one user put it, ``The [baseline feed] seemed to have listened to my preferences. I felt like the [interview] one didn't at all.'' 

Earlier, we raised the case of participants faulting the interview feed for not emphasizing the correct factors in the resulting feed ranking. This is a more direct piece of evidence that the interview is failing to elicit the relationship between preferences, even if it correctly identifies those different preferences. The majority of feed specifications, manually written or elicited by the interview, do not include explicit prioritization of different factors. However, the manual specifications, through their brevity, implicitly prioritize the content contained over that which is added by the interview. Therefore, when there is a clear hierarchy of preferences, the baseline may outperform the interview by prioritizing its few details in the ranking. When positivity is the main factor that matters, and the original specification is ``I'd like to see more positive posts to enjoy my life. I love reading posts helping me to grow and learn more. They can be about discovering new places and interesting ideas,'' of course that will outperform a feed ranked on 170 words about their relative preference for the inner workings of high-fashion ateliers.

Future iterations of this elicitation interview can mitigate this prioritization issue by more straightforwardly soliciting the importance of different user preferences. Following this study, we have implemented a recurring priority-ranking step after users give answers about their preferences. This prioritization data is incorporated into the feed specification, grouping the most important details together and explicitly calling out the only marginally important preferences as such, so they are not used to exclude potential posts, but to sort within already qualified content.  

\subsubsection{Prompts} \label{app: old prompts}

The following are the exact text of the prompts we used at various points of the interviewing, prompt-development, and post classification processes for the earlier iteration of our study. We only include prompts that differ between the two versions here.

\paragraph{Interviewer Prompt} \hfill\\
\texttt{You are a thoughtful, context-sensitive, reflective interviewer helping a person design a personalized social media feed that helps them to achieve a particular aim.
Craft your questions to gather deeper understanding and progress toward achieving the stated GOAL, while following the INTERVIEWING PRINCIPLES when crafting this question.}

\paragraph{Interviewing Principles} \hfill\\
\texttt{INTERVIEWING PRINCIPLES:
\begin{enumerate}
    \item Only ask one question at a time. Do not combine multiple into one. Do not ask questions that require more than a sentence or two to reply.
    \item Only ask questions that could be relevant to the textual contents of individual posts (not about the distribution of posts or post metadata like author)
    \item Provide examples or suggestions of things users may want to consider including or excluding beyond their existing specifications when relevant to the question.
    \item Don't force them to narrow down their topic area if they want broad content, especially if the topic is already niche. 
    \item  Aim for fewer than 4 questions per stage (ideal case is 2-3).
\end{enumerate}}

\begin{sloppypar}
\paragraph{Interview Goals for Stage 1} \hfill\\
\texttt{GOAL: The current stage of the interview is to figure out the purpose of the feed. The goal is to gather enough high-level information to clearly answer the following:
\begin{enumerate}
    \item Purpose or goal of the feed — What do they want to achieve with this feed? 
    \item Relevant user context — What is the user’s current situation, interests, or needs that this feed should address (if feed content relies on this information)?
    \item How social media can help with that goal — Broad framing, e.g., to inspire, expose, inform, entertain, make them feel something, etc.
    \item Desired emotional or experiential outcome — How the user wants to feel (or make others feel) during and after consuming the feed.
\end{enumerate}}    
\end{sloppypar}

\paragraph{Interview Goals for Stage 2} \hfill\\
\texttt{GOAL: The current stage of the interview is to identify the relevant topics and content for the feed to support the user goals. Ask questions in order to clearly understand the following:
\begin{enumerate}
    \item Desired topics/content — What additional related topics might be desired or relevant? Are there niche subtopics they want to focus on?
    \item Undesired but seemingly relevant topics/content — Topics or types of content that appear relevant but are actually unwanted or have a negative effect on the feed (distracting or upsetting).
\end{enumerate}}

\begin{sloppypar}
\paragraph{Interview Goals for Stage 3} \hfill\\
\texttt{GOAL: The current stage of the interview is to identify what qualities posts should have to support the user goals. Ask questions in order to clearly understand the following:
\begin{enumerate}
    \item Post quality attributes — What qualities or characteristics should be prioritized in posts (e.g., tone, depth, emotion, accuracy, creativity)?
    \item Quality exclusions — Are there qualities that should be avoided because they would harm the feed’s purpose or appeal?
\end{enumerate}}
\end{sloppypar}

\begin{sloppypar}
\paragraph{Interview Goals for Stage 4} \hfill\\
  \texttt{FOCUS: Ensure there is no remaining important information the user wants to share but has not yet had a chance to specify.}  
\end{sloppypar}

\paragraph{Stage 5 Summary Generation Prompt}  \hfill\\
\texttt{Synthesize their preferences into a plain text report of 2-3 paragraphs that explains what content they do or don’t want in their feed and how it should be ranked. Be clear about what is content is disallowed versus just not preferred (down-rank versus completely filter out) and organize by priority level. Avoid long lists of specific examples; instead, infer patterns, themes, and general rules from the provided information.}

\paragraph{Reflection Prompt} \hfill\\
\texttt{
Evaluate whether the listed GOAL has been met in the conversation so far.
A GOAL is met only if every related question specified in the GOAL has been answered with clear, 
complete, and sufficiently specific information, unless the user explicitly denies to answer. 
If the full answer to a question can be pieced together from multiple responses, consider that 
question answered.
If any answer is incomplete, unclear, vague, or requires further definition, consider that question 
not answered and the GOAL not met.
If the GOAL is met or the user has indicated a desire to proceed, output exactly: YES
If the GOAL is not met, output exactly: NO
Do not provide explanations or reasoning.}

\paragraph{Quality Classifier Creation Prompt} \hfill\\
\texttt{If it is not already in that form, synthesize the preferences into a 2-3 plain text paragraph report that explains what content they do or don't want in their feed and how it should be ranked. DO NOT EXTRAPOLATE. Focus ONLY on what is explicitly said in the description.
}


\end{document}